\documentclass[aps,prb,twocolumn,groupedaddress,superscriptaddress,footinbib]{revtex4-2}

\usepackage{amsmath}
\usepackage[dvips]{color}
\usepackage[utf8]{inputenc}
\usepackage[american]{babel}
\usepackage{color}
\usepackage{graphicx}
\usepackage{epstopdf}
\usepackage{epsfig}
\usepackage{float}
\usepackage[colorlinks,linkcolor=blue,citecolor=blue,urlcolor=blue]{hyperref}
\usepackage{comment}
\usepackage{amssymb}
\usepackage{multirow}
\usepackage[table]{xcolor}
\usepackage[normalem]{ulem}

\allowdisplaybreaks

\usepackage[normalem]{ulem}
\usepackage{orcidlink}

\begin{document}

\newcommand*{\TUW}[0]{{Institute of Solid State Physics, TU Wien, 1040 Vienna, Austria}}

\newcommand*{\KACIF}[0]{{Department for Research of Materials under Extreme Conditions, Institute of Physics, 10000 Zagreb, Croatia}}

\newcommand*{\CALT}[0]{{Centre for Advanced Laser Techniques, Institute of Physics, 10000 Zagreb, Croatia}}

\newcommand*{\COLOGNE}[0]{{Institute for Theoretical Physics, University of Cologne, 50937 Cologne, Germany}}

\newcommand*{\FLATIRON}[0]{{Center for Computational Quantum Physics, Flatiron Institute, 162 5th Avenue, New York, New York 10010, USA}}

\author{Juraj Krsnik\orcidlink{0000-0002-4357-2629}}
\email{jkrsnik@ifs.hr}
\affiliation{\KACIF}

\author{Dino Novko\orcidlink{0000-0002-1673-2973}}
\email{dnovko@ifs.hr}
\affiliation{\CALT}

\author{Fabian B. Kugler\orcidlink{0000-0002-3108-6607}}
\affiliation{\COLOGNE}

\author{Osor S. Bari\v{s}i\'c\orcidlink{0000-0002-6514-9004}}
\affiliation{\KACIF}

\author{Karsten Held\orcidlink{0000-0001-5984-8549}}
\affiliation{\TUW}





\title{Unconventional plasmon dynamics due to strong correlations in Sr$_2$RuO$_4$}


\begin{abstract}
	
Plasmon modes,
their dispersion, and the onset of 
damping when approaching the electron-hole continuum are well understood when electron correlations are weak. However, we know little about how this picture is modified and what additional features emerge in strongly correlated materials.
Here, we present a fully \textit{ab initio} approach
to plasmon excitations that combines density functional theory with dynamical mean-field theory, and we use it to reconcile controversial electron energy-loss spectroscopy results in Sr$_2$RuO$_4$.
In particular, we show that 
electronic correlations reproduce the 
plasmon dispersion, while generating a large intrinsic width already below the electron-hole continuum. An additional high-energy peak reflecting transitions between incoherent features and a sharp increase of the plasmon’s energy-momentum dispersion, akin to waterfalls in photoemission spectroscopy, are identified as genuine correlation effects.
	
\end{abstract}

\maketitle 

Collective charge excitations -- plasmons -- of correlated systems such as ruthenates, cuprates, and nickelates provide important insight into the role of electron dynamics in  phenomena such as unconventional high-temperature superconductivity and peculiar transport properties\,\cite{basov2011}. 
For instance, plasmons are discussed as a potential mediator of electron pairing\,\cite{ruvalds1987,kresin1988,nag2020} and as a probe for correlated states\,\cite{papaj2023}.
However, the nature of the plasmonic response in these materials is a highly controversial topic\,\cite{abbamonte2025,husain2019,fink2021,husain2021}, where some authors report that plasmons are overdamped well below the Landau damping region\,\cite{bozovic1987,mitrano2018,romero2019,devries2026}, while others discuss plasmon excitations as unrenormalized modes\,\cite{nucker1989,knupfer2022,Francesco2025}. These results suggest that some aspects of plasmon physics are (deceivingly or not)  well described within the 
textbook random-phase approximation (RPA), 
while other segments assumably involve strong correlations.

Recent electron energy-loss spectroscopy (EELS) experiments\,\cite{knupfer2022,husain2023,schultz2025} have revealed a particularly curious plasmonic response in the layered perovskite Sr$_2$RuO$_4$, a correlated material that exhibits unconventional superconductivity below $T_c\sim 1$\,K\,\cite{maeno1994,mackenzie1998} and a linear-in-temperature resistivity above $\sim$ 600\,K\,\cite{tyler1998}. 
Transmission EELS measurements report a conventional (RPA-like) mode 
that starts around $\omega_{pl}\sim 1.6$\,eV in the long-wavelength region, in agreement with the optical reflectivity data\,\cite{katsufuji1996,hildebrand2001,stricker2014}, entering a Landau continuum at $q_c=0.4$\,\AA$^{-1}$. 
However, EELS also shows several puzzling unconventional features that are hitherto not understood:
(1) While the plasmon dispersion agrees with RPA\;\cite{knupfer2022,schultz2025},
its width $\Gamma_{pl}\sim 1$\,eV\,\cite{knupfer2022,schultz2025} is huge, completely out of bounds of  RPA. (2) A second broad peak of unknown origin is observed at $\sim 3.3$\,eV\,\cite{knupfer2022}. (3) Reflection EELS shows a broad continuum or overdamped plasmon in the large momentum range peaked at $\sim 1.2$\,eV, with a slightly negative dispersion\,\cite{husain2023}. These results are reminiscent of the puzzling results in cuprates and thus point to a general, unresolved problem of plasmon dynamics in strongly correlated materials\,\cite{abbamonte2025}. In addition, at energies of about $\sim 10$\,meV another peak was reported in Sr$_2$RuO$_4$\,\cite{husain2023}, interpreted as an acoustic ``demon'' mode \cite{pines1956,Ruvalds1981,sanchez2026}. 

So far, plasmons in Sr$_2$RuO$_4$ have been theoretically modeled only with  RPA-like methods, using a tight-binding approximation\,\cite{husain2023,choi2024} or density functional theory (DFT)\,\cite{hochberg2025}  or  the Fetter multilayer expression\,\cite{fetter1974,apostol1975,knupfer2022,schultz2025,devries2025} as a starting point. The corresponding simulations show a highly anisotropic response, where for the out-of-plane momentum $q_\perp=0$  electrons in all layers oscillate in phase forming a bulk optical plasmon, while as $q_\bot$ increases, the plasmon energy at 
the in-plane momentum $|\mathbf{q}_{\Vert}|=0$ decreases and shows an acoustic-like dispersion. However, these previous theoretical results are unable to capture the aforementioned unconventional features of plasmons in Sr$_2$RuO$_4$, which is not surprising considering that photoemission and optical measurements report strong renormalizations 
of single-particle properties\,\cite{ingle2005,Iwasawa2012,tamai2019,Hunter2023,katsufuji1996,stricker2014}. This calls for a thorough theoretical study of plasmons with a realistic description of its electronic structure and electron correlations, beyond a RPA picture or model Hamiltonians \cite{khaliullin1996,Loon2014,Hafermann2014,greco2016,Philoxene2024,Yamase2026,yamase2026_2}.

In this letter, we utilize a fully \textit{ab initio} combination of DFT and dynamical mean-field theory (DMFT) \cite{Georges1996,Kotliar2006}, which accounts for multiorbital electronic structure, 
spin-orbit coupling (SOC), and electron correlations via a dynamical self-energy that is local in the orbital basis, to perform a detailed study of plasmon excitations in Sr$_2$RuO$_4$. Our analyses reveal rich excitation spectra, reconciling seemingly contradictory EELS data and rationalizing an RPA-like optical plasmon energy\,\cite{schultz2025,husain2023,abbamonte2025}: we obtain an optical plasmon pole at $\sim 1.5$\,eV around $|\mathbf{q}_{\Vert}|=0$ with large electron-correlation-induced width of $\sim1$\,eV, a
plasmon gap of $\sim20$ meV for $q_\bot>0$, a second high-energy peak due to correlation-driven features at $\sim 3$\,eV, negative dispersion at larger $|\mathbf{q}_{\Vert}|$, as well as the low-energy peak around $\sim$ 10 meV. The latter is, however, not a well-defined pole.
In addition, we argue that strong renormalizations of a single-particle spectrum translate into a sharp increase of the plasmon's energy-momentum dispersion, akin to waterfalls \cite{si2024,krsnik2025} in angle-resolved photoemission spectroscopy (ARPES) \cite{Valla2007,Xie2007,Meevasana2007,Iwasawa2010,Iwasawa2012,ding2024,Sun2025}. A sharp increase of the plasmon dispersion in some hole-doped cuprates \cite{singh2022,nag2020,Nakata2025} is reminiscent of this plasmon waterfall effect. However, this interpretation calls for a more thorough future analysis.

 \begin{figure}
 	\includegraphics[width=0.485\textwidth]{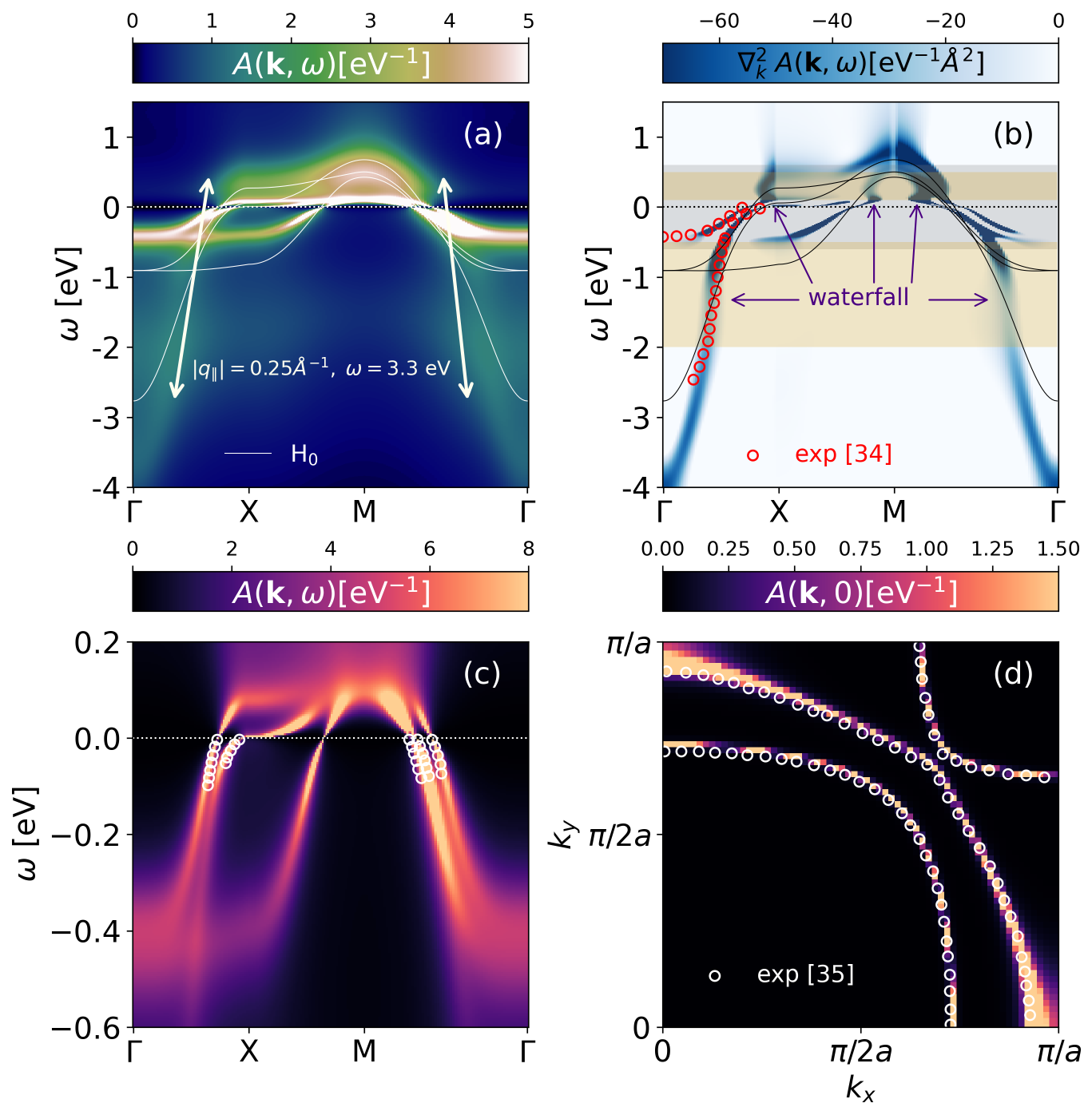}
 	\renewcommand{\figurename}{Fig.}
 	\caption{DFT+DMFT (a,c) spectral function, (b) its second derivative with respect to momentum, and (d) Fermi surface for Sr$_2$RuO$_4$  together with the DFT-derived $t_{2g}$ (solid white/black lines) band dispersions compared to ARPES [red~\cite{Iwasawa2012} and white~\cite{tamai2019} circles in panels (b) and (c,d), respectively]. White arrows in (a) denote the energy and momentum transfer at which the high-energy peak is first observed experimentally \cite{knupfer2022}. The gray shaded area in (b), corresponding to $\omega\in [-0.6,0.6]$ eV, indicates the energy range contributing to the layered plasmon dispersion, where it exhibits a steep increase, while the gold shaded areas highlight the ranges where the waterfall features appear, indicated by the purple arrows.}
 	\label{fig:fs_sf_sd}
 \end{figure}
 
  \begin{figure*}
 	\includegraphics[width=1\textwidth]{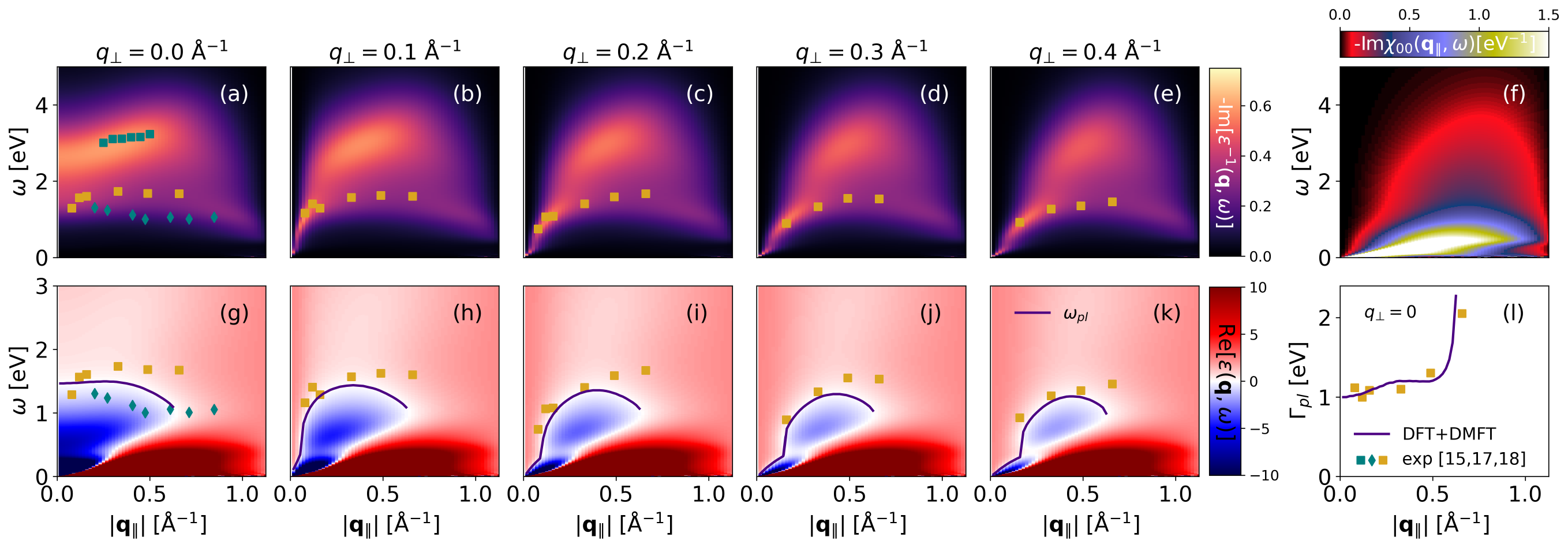}
 	\renewcommand{\figurename}{Fig.}
 	\caption{DFT+DMFT (a-e) EELS spectra $\text{Im}\left[ \varepsilon(\mathbf{q},\omega)\right]^{-1}$,  and (g-k) Re$\left[ \varepsilon(\mathbf{q},\omega)\right] $ for five out-of-plane momenta $q_\perp$ and in-plane momentum $\mathbf{q}_\Vert$ along the [11] direction, together with the measured EELS peaks (teal squares \cite{knupfer2022}, teal diamonds \cite{husain2023}, golden squares \cite{schultz2025})  and DFT+DMFT (purple lines) plasmon dispersions. (f)  Imaginary part of the DFT+DMFT $\chi_{00}(\mathbf{q}_\Vert,\omega)$ depicting the electron-hole continuum. (l) Measured (golden squares) \cite{schultz2025} and DFT+DMFT (purple line) plasmon half width at half maximum for $q_\bot=0$.}
 	\label{fig:eels_re_epsilon_gamma}
 \end{figure*}
 
The DFT calculations were performed using \verb|Quantum ESPRESSO| \cite{Giannozzi_2017} with DFT-relaxed lattice parameters $a=b=3.88\;\textup{\AA}$ and $c=12.86\;\textup{\AA}$.
Optimized norm-conserving Vanderbilt scalar-relativistic pseudopotentials constructed with the  Perdew–Burkef–Ernzerhof~(PBE) exchange-correlation functional were employed \cite{vanSetten2018}. SOC was not taken into account at the DFT level. Following well established results for the electronic structure of Sr$_2$RuO$_4$ \cite{zhang2016,tamai2019}, we further generated a Hamiltonian $\mathbf{H}_{0}(\mathbf{k})$ with maximally localized Wannier functions for the three bands crossing the Fermi level.  These are constructed from the three ruthenium $4d$ $(d_{xy}, d_{xz}, d_{yz})$ orbitals using \verb|wannier90| \cite{mostofi2014}, yielding three bands shown as solid white/black lines in Fig.~\ref{fig:fs_sf_sd}(a)/(b).

To correctly reproduce the experimentally measured low-energy electronic structure\,\cite{Iwasawa2010,Iwasawa2012,tamai2019}, SOC and electron correlations need to be included. Following Refs.~\cite{zhang2016,tamai2019,Suzuki2023}, we use the DMFT  $t_{2g}$ self-energy $\boldsymbol{\Sigma}(\omega)$ without SOC, subtract its static part $\boldsymbol{\Sigma}(\omega)\to\boldsymbol{\Sigma}(\omega)-\text{Re}\boldsymbol{\Sigma}(\omega=0)$, and subsequently add a local SOC $\mathbf{H}_{\rm SOC}=\frac{\lambda}{2}\mathbf{l}\cdot{{\sigma}}$
with $\lambda=0.2$\,eV enhanced by electron correlations. 
Here, $\mathbf{l}$ are the 
angular-momentum matrices in the $t_{2g}$-basis  and ${{\sigma}}$ are Pauli matrices (we only denote $t_{2g}$ matrices in bold). We calculate the retarded $\boldsymbol{\Sigma}(\omega)$ directly on the real-frequency axis using the numerical renormalization group (NRG) \cite{Wilson1975,Bulla2008} at $T=5$K and $(U,J)=(2.3,0.4)\,\mathrm{eV}$ \cite{Kugler2020}. 
Specifically, $\boldsymbol{\Sigma}(\omega)$ was obtained in an implementation~\cite{Kugler2022a,Kugler2024,Grundner2025,Kugler2026,LaBollita2026} using the MuNRG package \cite{Lee2016,Lee2017,Lee2021,Kugler2022} built on top of the QSpace tensor library \cite{Weichselbaum2012a, Weichselbaum2020, Weichselbaum2024}. 
Thus, our calculations avoid analytic continuation. 
The chemical potential $\mu$ is adjusted by imposing that the filling is equal to four
electrons per ruthenium ion \cite{tamai2019}. The trace of the resulting spectral function
$\mathbf{A}(\mathbf{k},\omega)=-\pi^{-1}\text{Im} \left[\left( \omega+\mu\right)\cdot\mathbf{I} -\mathbf{H}_{0}(\mathbf{k})-\mathbf{H}_{\rm SOC}-\boldsymbol{\Sigma}(\omega) \right] ^{-1},
$
shown in Fig.~\ref{fig:fs_sf_sd}, is in excellent agreement with experiments \cite{Iwasawa2012,tamai2019} (red and white circles).

Having specified the \emph{ab initio} model Hamiltonian, we first evaluate the retarded current-dipole correlation function $\chi_{\alpha\beta}(\mathbf{q},\omega)$
from the DFT+DMFT spectral function, neglecting vertex corrections. Similar as for the DMFT optical conductivity \cite{khurana1990}, vertex corrections to $\chi_{\alpha\beta}(\mathbf{q},\omega)$ vanish for $\mathbf{q}\rightarrow 0$ and a single band, and can thus be expected to be small for small $\mathbf{q}$ and the three $t_{2g}$ orbitals.
Employing the continuity equation \cite{kubo2012statistical,schrieffer2018,Kupcic2013,novko2016}, the density-density correlation function is then obtained as $\chi_{00}(\mathbf{q,\omega})=\frac{\hbar}{i\omega}\sum_{\alpha\beta}q_\alpha\chi_{\alpha\beta}q_\beta$. Our $\omega$ has units of energy.

To describe the  experimentally observed long-range plasmons we further include the non-local Coulomb interaction $V({\mathbf q})$, which does not affect the DMFT self-energy (as long as there is no ordering such non-local interactions reduce to the Hartree term \cite{MullerHartmann1989a}).
However, together with the DFT+DMFT-calculated charge response, $V({\mathbf q})$ determines the dielectric function in the leading monopole-monopole approximation $\varepsilon(\mathbf{q},\omega)=\varepsilon_\infty - V({\mathbf{q}})\chi_{00}(\mathbf{q,\omega})$ \cite{Kupcic2014,zupanovic1997_1,kupcic2000}. For a layered system, $V(\mathbf{q}_\Vert,q_\bot)=\frac{e^2}{2\epsilon_0 |\mathbf{q}_{\Vert}|A}\frac{\sinh\left( |\mathbf{q}_{\Vert}|d \right) }{\cosh( |\mathbf{q}_{\Vert}|)-\cos(q_\bot d)}$  \cite{fetter1974,schultz2025,shen2026}, where $A=a^2$ and $d=c/2$. We take $\varepsilon_\infty\approx2.3$ from experiment \cite{stricker2014} as it stems from processes beyond our three-band DFT-derived Hamiltonian.
This way of combining DMFT and the monopole-monopole approximation allows us to finally evaluate the loss function $\text{Im}\left[ \varepsilon(\mathbf{q},\omega)\right]^{-1}$. Owing to the strong anisotropy of Sr$_2$RuO$_4$ \cite{beck2025}, we model the response functions as effectively two-dimensional,  $\chi(\mathbf{q},\omega)\to \chi(\mathbf{q}_\parallel,\omega)$. Consequently, the out-of-plane momentum dependence of the dielectric function enters solely through Coulomb interaction.

This DFT+DMFT approach, which evaluates the charge response from the current-dipole correlation function and continuity equation, satisfies the optical sum rule to a great extent. It has proven to be particularly accurate in describing the plasmon energy and lifetime within the relaxation-time approximation \cite{Kupcic2014,Kupcic2015}, as well as the electron-phonon memory function \cite{novko2017}. Further details on the modeling and the optical sum rule are provided in Secs.~S1 and S2 in the Supplemental Material (SM) \cite{SM}.

Reference~\cite{schultz2025} reported measurements of the energy-momentum dispersion of the plasmon in Sr$_2$RuO$_4$ for five out-of-plane momenta, $q_\perp = 0$, $0.1$, $0.2$, $0.3$, and $0.4~\text{\AA}^{-1}$. The corresponding dispersions are shown by golden squares in Figs.~\ref{fig:eels_re_epsilon_gamma}(a-e), on top of our DFT+DMFT EELS spectra for in-plane momenta $\mathbf{q}_\Vert$ along the [11] direction. Our calculated spectra exhibit a two-peak structure, where the peak at lower energies follows the measured plasmon dispersion\,\cite{schultz2025}. To determine the plasmon energy, we note that longitudinal plasma oscillations manifest as zeros of the real part of the dielectric function, shown in Figs.~\ref{fig:eels_re_epsilon_gamma}(g-k). For each momentum, we observe only a single zero crossing from negative to positive values of $\text {Re}\left[\varepsilon(\mathbf{q},\omega)\right] $ with increasing $\omega$, defining the plasma energy via $\text{Re}\left[ \varepsilon(\mathbf{q},\omega_{pl}(\mathbf{q}))\right] =0$, shown in Figs.~\ref{fig:eels_re_epsilon_gamma}(g-k) as a purple line. The fully \textit{ab initio} DFT+DMFT optical plasmon energy, $\omega_{pl}(|\mathbf{q}|\to0)\approx1.47$ eV, is in excellent agreement with experimental values \cite{knupfer2022,husain2023,schultz2025}.
This agreement persists at finite momenta. Only for $|\mathbf{q}_\Vert| > 0.5~\text{\AA}^{-1}$ our calculated plasmon energy starts to be underestimated compared to transmission EELS results\,\cite{schultz2025}. As seen in Fig.~\ref{fig:eels_re_epsilon_gamma}(f), which shows $\mathrm{Im}\,\chi_{00}(\mathbf{q},\omega)$, at these large momenta the plasmon dispersion approaches the electron-hole continuum and eventually ceases to represent a well-defined excitation [$\mathrm{Im}\,\chi_{00}$ is small (black region) only for small $|\mathbf{q}_\parallel|$]. This agrees with experimental EELS peaks that become strongly broadened for large in-plane momenta $q_c > 0.4~\text{\AA}^{-1}$ \cite{schultz2025}. In fact, this large-momentum region is more aligned with the reports of reflection EELS results\,\cite{husain2023}, that show a broad mode at $\sim$1.2\,eV and a slightly negative dispersion [teal diamonds in Figs.~\ref{fig:eels_re_epsilon_gamma}(a,g)]. As discussed in more detail in Sec.~S3 of the SM \cite{SM}, such a peculiar plasmonic response is not governed by a single factor but emerges from the combined effect of the optical effective mass, spectral weight, screening, and damping. Note that, since reflection EELS is more surface sensitive, the experimentally observed lower plasmon energy may arise from contributions of a surface plasmon, which typically has a lower energy than a bulk plasmon \cite{Pitarke2007}.

\begin{figure}
	\includegraphics[width=0.49\textwidth]{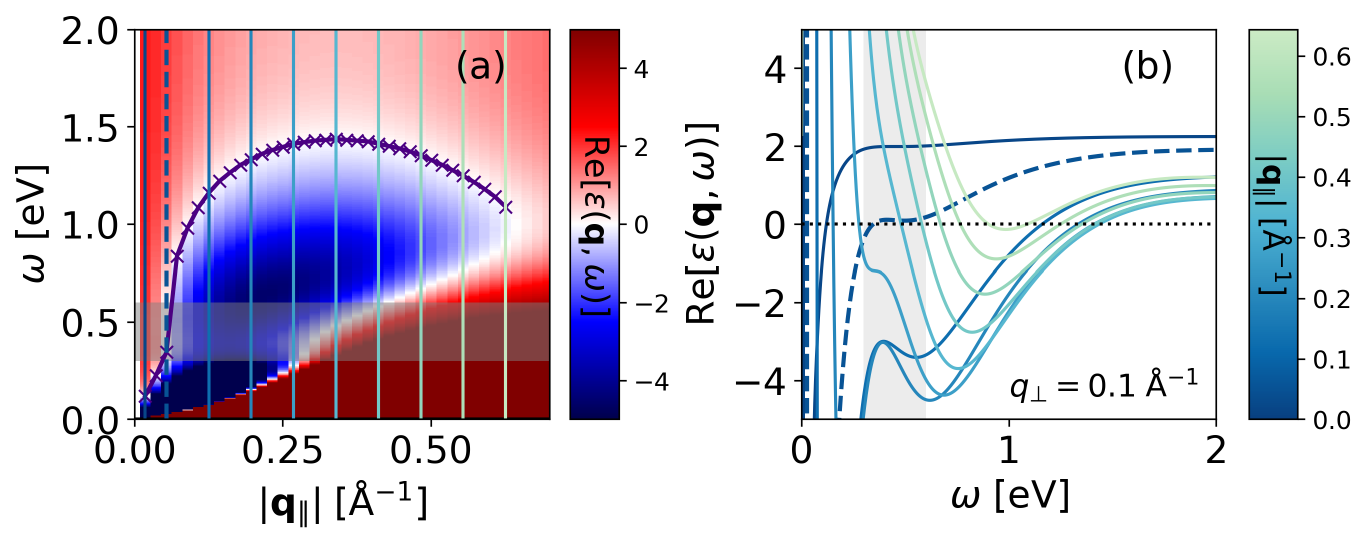}
	\renewcommand{\figurename}{Fig.}
	\caption{(a) Real part of the DFT+DMFT dielectric function, Re$\left[ \varepsilon(\mathbf{q},\omega)\right] $, for the out-of-plane momentum $q_\perp=0.1~\text{\AA}^{-1}$. Colored vertical lines denote the in-plane momenta $|\mathbf{q}_\Vert|$ for which the energy dependence of  Re$\left[ \varepsilon(\mathbf{q},\omega)\right] $ is shown in panel (b). The dashed line denotes the momentum at which the plasmon dispersion (purple line) exhibits a steep increase due to Re$\left[ \varepsilon(\mathbf{q},\omega)\right]$ being close to 0 (dotted black line in (b)) over a continuous energy range (shaded area).}
	\label{fig:plasmon_waterfall}
\end{figure}

\begin{figure*}
	\includegraphics[width=1\textwidth]{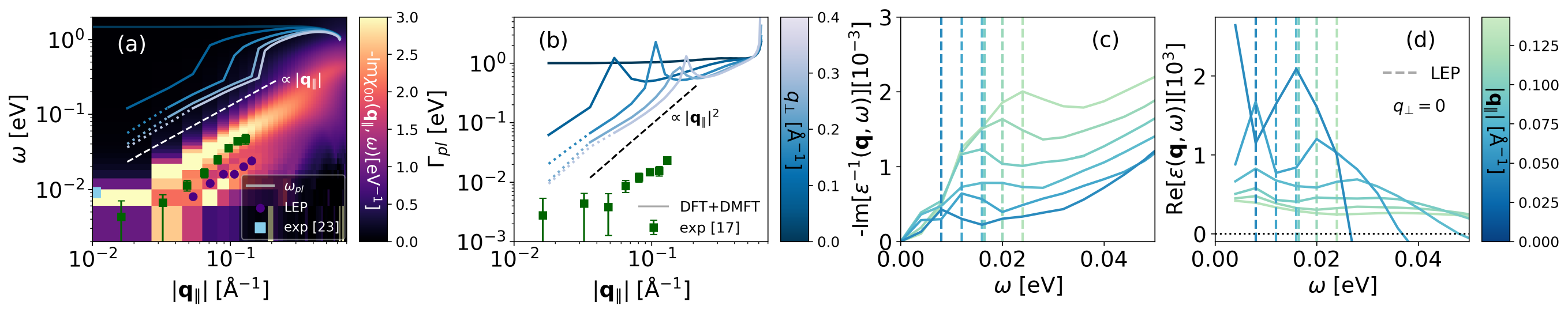}
	\renewcommand{\figurename}{Fig.}
	\caption{(a) False-color plot of the imaginary part of the DFT+DMFT $\chi_{00}(\mathbf{q}_\Vert,\omega)$, depicting the electron-hole continuum, together with the plasmon dispersions for different $q_\bot$ added on top (colored lines; the corresponding color code is next to panel (b)). 
    The blue square denotes the resonance energy from optical spectroscopy \cite{hildebrand2001}.
    (b) Plasmon half width at half maximum for different out-of-plane momenta $q_\bot$, shown on a log-log scale. The $|\mathbf{q}_\Vert|$ and $|\mathbf{q}_\Vert|^2$ dependencies in (a) and (b), respectively, are guides to the eye. (c) Low-energy DFT+DMFT EELS spectra and (d) real part of the dielectric function (note the scaling factors) for $q_\bot=0$ and several momenta $|\mathbf{q}_\Vert|$. Together, they reveal a strongly broadened low-energy peak (LEP), indicated by vertical dashed lines, whose dispersion is shown with purple circles in (a). The latter shows good agreement with the peak positions extracted from EELS measurements for $q_\bot=0$ \cite{husain2023} (green squares in (a)), attributed to an acoustic ``demon'' mode. The measured width of the corresponding low-energy peak is shown in (b) (green squares).}
	\label{fig:low_energy}
\end{figure*}

The high-energy peak, also observed in experiments near $3.3$ eV (teal squares \cite{knupfer2022,schultz2025} in Fig.~\ref{fig:eels_re_epsilon_gamma}), while showing some hybridization with the plasmon, should not be identified with the plasmon itself. In particular, this feature is not present in the RPA modeling \cite{knupfer2022,husain2023,schultz2025}. In fact, the $\sim 3$\,eV-peak lies within the interband gap in our RPA calculations based on the full DFT electronic structure (see Fig.~S5 in the SM\,\cite{SM}). This DFT calculation includes high-energy interband transitions beyond the three-band model, but {\it not} the DMFT self-energy.

Fig.~\ref{fig:fs_sf_sd}(a) highlights possible transitions in the single-particle spectral function at the energy and momentum transfer where the high-energy peak  at 3.3\,eV is first observed experimentally. These transitions apparently involve incoherent spectral features, further supporting the interpretation of this experimental peak as a correlation-driven effect \cite{Loon2014,basov2011,ayral2012}.


By expanding the loss function around the plasmon pole, the plasmon half width at half maximum can be estimated as

\begin{equation} \label{eq:plasmon_width}
	\Gamma_{pl}(\mathbf{q})= \frac{\text{Im}\,\varepsilon(\mathbf{q},\omega_{pl}(\mathbf{q}))}{\left.\frac{\partial \text{Re}\,\varepsilon(\mathbf{q},\omega)}{\partial \omega}\right|_{\omega=\omega_{pl}(\mathbf{q})}}\;.
\end{equation}
The thus calculated  DFT+DMFT in-plane momentum dependence of $\Gamma_{pl}(|\mathbf{q}_\Vert|,q_\bot=0)$ (purple line), is plotted together with the experimental values (golden squares) \cite{schultz2025} for $q_\bot = 0~\text{\AA}^{-1}$ in Fig.~\ref{fig:eels_re_epsilon_gamma}(l). While the  RPA captures only {\it qualitatively} the increase of the plasmon width due to Landau damping \cite{schultz2025} (see Fig.~S4(b) in the SM \cite{SM}), our DFT+DMFT results also show excellent {\it quantitative} agreement with experiment. In the long-wavelength limit, the calculated plasmon width correctly reproduces the experimental value $\Gamma_{pl}(|\mathbf{q}|\to0)\approx 1$ eV. It remains nearly constant up to the vicinity of the electron-hole continuum, where it increase to $\sim$ 2 eV. These results indicate that electron correlations captured within DMFT are both essential and sufficient for an accurate description of the plasmon width in Sr$_2$RuO$_4$.

Another signature of strong electron correlations in our calculated plasmon properties is the sharp increase in the plasmon energy-momentum dispersion observed in Figs.~\ref{fig:eels_re_epsilon_gamma}(h-k) within the energy interval $\sim0.3-0.6$ eV. To gain deeper insight into this behavior, Fig.~\ref{fig:plasmon_waterfall}(a)
takes a closer look at $\text {Re}\left[\varepsilon(\mathbf{q},\omega)\right]$ for the out-of-plane momentum $q_\perp=0.1~\text{\AA}^{-1}$ along with  Fig.~\ref{fig:plasmon_waterfall}(b), where we plot the energy dependence of $\text {Re}\left[\varepsilon(\mathbf{q},\omega)\right] $ for the in-plane momenta denoted by the colored vertical lines in Fig.~\ref{fig:plasmon_waterfall}(a). Around the momentum denoted by the dashed line, the plasmon dispersion (purple line) exhibits a steep increase. This behavior of the collective dynamics can be traced to the condition $\text {Re}\left[\varepsilon(\mathbf{q},\omega_{pl}(\mathbf{q}))\right] \approx 0$ being satisfied over a continuous energy range (shaded area) for a single $\mathbf{q}$ point.

Such a feature is reminiscent of waterfalls observed in ARPES spectra \cite{Valla2007,Xie2007,Meevasana2007,ding2024,Sun2025}, where the pole equation is fulfilled across a continuous range of energies \cite{si2024,krsnik2025}, giving rise to a waterfall-like feature connecting the quasiparticle band and Hubbard bands. In fact, waterfalls have been reported in ARPES spectra of Sr$_2$RuO$_4$ starting at around $\sim 0.5$ eV [red circles in Fig.~\ref{fig:fs_sf_sd}(b)] \cite{Iwasawa2010,Iwasawa2012} and are also present in our DFT+DMFT spectrum [indicated by the purple arrows within the gold shaded areas in Fig.~\ref{fig:fs_sf_sd}(b)] over a similar energy range. On the other hand, the gray shaded region in Fig.~\ref{fig:fs_sf_sd}(b) marks the energy range in which the plasmon dispersion exhibits a steep waterfall-like increase. The latter energy range is spanned by transitions between the coherent single-particle spectral weight around the Fermi level and the waterfall-like spectral weight above the Fermi level, suggesting that the steep plasmon dispersion originates from electron correlations. Indeed, such a plasmon dispersion is absent in RPA calculations \cite{knupfer2022,husain2023,schultz2025}, indicating that it should not be attributed to specific details of the non-interacting band structure \cite{silkin2023,Silkin2025}.

Experimentally the plasmon waterfall effect in Sr$_2$RuO$_4$ can be tested by probing the plasmon dispersion at smaller in-plane momenta than in Ref.~\cite{schultz2025}. We anticipate similar effects for gapless and layered plasmons in other effectively two-dimensional correlated materials\,\cite{nag2020,Hepting2022,bejas2024,nag2024,shen2026,krsnik2024_sc,Silkin2025,devries2025,devries2026,zinni2026} as well, where the sharp dispersion upturn accompanied by enhanced plasmon broadening reported in hole-doped cuprates \cite{singh2022} appear as a promising candidate for such correlation-driven behavior. In particular, following Eq.~\eqref{eq:plasmon_width}, the steep increase of the plasmon energy is accompanied by a pronounced enhancement of the plasmon width, see Fig.~\ref{fig:low_energy}(b), again analogous to the strongly broadened waterfalls in ARPES.

At the smallest momenta considered in our calculations, the DFT+DMFT plasmon dispersion becomes linear, acoustic-like in $|\mathbf{q}_\Vert|$ for finite out-of-plane momenta $q_\bot$, $\omega_{pl}(\mathbf{q}_\Vert, q_\bot>0)\approx \hbar v_{pl}(q_\bot)|\mathbf{q}_\Vert|$, see the colored lines in Fig.~\ref{fig:low_energy}(a). Linear fits yield $\hbar v_{pl}(q_\bot)\approx 6.39$, $3.27$, $2.35$, and $2$ eV$\text{\AA}$, for $q_\perp = 0.1$, $0.2$, $0.3$, and $0.4~\text{\AA}^{-1}$, respectively.  The corresponding plasmon widths scale as $|\mathbf{q}_\Vert|^2$, as demonstrated in Fig.~\ref{fig:low_energy}(b). Following the discussion in Sec.~S3 of the SM \cite{SM}, the layered plasmon, however, crosses over from a linear to a quadratic dispersion as $|\mathbf{q}_\Vert|\to0$ [dotted lines in Fig.~\ref{fig:low_energy}(a) indicate the portions of the dispersion that should deviate from linear behavior] and develops a small but finite gap \cite{greco2016,Hepting2022} due to the finite out-of-plane hopping. We estimate the latter to be of order $20$ meV. Optical spectroscopy revealed a well-defined resonance at 9\,meV\,\cite{katsufuji1996,hildebrand2001} [blue square in Fig.~\ref{fig:low_energy}(a)], which might be attributed to this mode.

Reference~\cite{husain2023} also reported a low-energy peak in EELS, however for $q_\bot=0$, ascribed to the acoustic ``demon'' mode \cite{pines1956,Ruvalds1981}, whose positions and widths are shown by green squares in Figs.~\ref{fig:low_energy}(a) and (b), respectively. The peak dispersion is linear over most of the measured energy-momentum range but behaves quadratic at small momenta. A possible gap is hindered by the elastic peak. While this phenomenology qualitatively resembles the finite-$q_\bot$ layered plasmon dispersion discussed above, the characteristic energy scales differ by nearly an order of magnitude. Interestingly though, the measured peak positions lie between the two maxima of the DFT+DMFT electron-hole continuum [false-color plot in Fig.~\ref{fig:low_energy}(a)], a condition favorable for the emergence of a multiorbital acoustic plasmon \cite{pines1956,Ruvalds1981,silkin2023,Silkin2025,Pisarra2014}. A closer inspection of our loss function for $q_\bot=0$ [Fig.~\ref{fig:low_energy}(c)] indeed reveals a small low-energy peak with a dispersion [purple circles in Fig.~\ref{fig:low_energy}(a)] in reasonable agreement with the low-energy peak in EELS, i.e., the peak ascribed to the acoustic demon mode. However, although $\text{Re}\,\varepsilon(\mathbf{q},\omega)$ develops a shallow local minimum [Fig.~\ref{fig:low_energy}(d)], it does not cross zero and is accompanied by strong damping (see also Sec.~S5 in the SM\,\cite{SM}). This feature, of predominantly intraband origin, is therefore not a genuine pole.

To conclude, building on the successful description of single-particle properties in Sr$_2$RuO$_4$ based on strong electron correlations, we demonstrate that the collective charge dynamics can also be accurately captured within this framework, provided that spectral weight, optical effective mass, screening, and damping are all properly accounted for. Only then,  excellent agreement with  EELS and optical measurements is achieved. Electronic correlations naturally explain the enormous damping of the plasmon mode, a second high-energy peak at 3.3\,eV, and predict a sudden increase of the plasmon dispersion
at $0.3-0.6$\,eV. Correlations also renormalize the peculiar ``demon'' feature, which is already present at the DFT level; however, in neither case does it constitute a well-defined mode. 

\begin{acknowledgments}
The authors thank Luke C. Rhodes for useful comments. J.K., D.N., and O.S.B.\ acknowledge financial support from the Croatian Science Foundation (Grant nos. UIP-2025-02-5952, IP-2025-02-5926, and IP-2022-10-3382, respectively). J.K. and O.S.B., and D.N. acknowledge financial support from the project FrustKor and from the project ``Podizanje znanstvene izvrsnosti Centra za napredne laserske tehnike (CALTboost)", respectively, financed by the EU through the National Recovery and Resilience Plan 2021-2026 (NRPP).
F.B.K.\ acknowledges funding from the Ministerium f\"ur Kultur und Wissenschaft des Landes Nordrhein-Westfalen (NRW-R\"uckkehrprogramm).
F.B.K.\ and K.H.\ acknowledge support from the Deutsche Forschungsgemeinschaft (DFG) research unit
FOR5249 (``QUAST'', project no.~449872909) and the Austrian science funds (FWF) QUAST (sub)project DOI 10.55776/KIN2563725, respectively. J.K.\ and K.H.\ have been supported by the FWF project DOI 10.55776/P36213.
\end{acknowledgments}

\bibliography{bibliography}

\end{document}


\title{Supplemental Material:\\ Unconventional plasmon dynamics due to strong correlations in Sr$_2$RuO$_4$}

\newcommand*{\TUW}[0]{{Institute of Solid State Physics, TU Wien, 1040 Vienna, Austria}}

\newcommand*{\KACIF}[0]{{Department for Research of Materials under Extreme Conditions, Institute of Physics, 10000 Zagreb, Croatia}}

\newcommand*{\CALT}[0]{{Centre for Advanced Laser Techniques, Institute of Physics, 10000 Zagreb, Croatia}}

\newcommand*{\COLOGNE}[0]{{Institute for Theoretical Physics, University of Cologne, 50937 Cologne, Germany}}

\newcommand*{\FLATIRON}[0]{{Center for Computational Quantum Physics, Flatiron Institute, 162 5th Avenue, New York, New York 10010, USA}}

\author{Juraj Krsnik\orcidlink{0000-0002-4357-2629}}
\email{jkrsnik@ifs.hr}
\affiliation{\KACIF}

\author{Dino Novko\orcidlink{0000-0002-1673-2973}}
\email{dnovko@ifs.hr}
\affiliation{\CALT}

\author{Fabian B. Kugler\orcidlink{0000-0002-3108-6607}}
\affiliation{\COLOGNE}

\author{Osor S. Bari\v{s}i\'c\orcidlink{0000-0002-6514-9004}}
\affiliation{\KACIF}

\author{Karsten Held\orcidlink{0000-0001-5984-8549}}
\affiliation{\TUW}

\maketitle 

\section{Current-dipole correlation function}

Our density functional theory (DFT)-derived Hamiltonian
$\hat{\mathbf{H}}_{\text{0}}$ describing the 
low-energy physics in Sr$_2$RuO$_4$ consists of three maximally localized Wannier orbitals constructed from the three ruthenium  $4d$ $t_{2g}$ orbitals. Together with electron spin, this forms a set of six electron states $\left\lbrace n\right\rbrace $, with accompanying wave functions $\phi_n$, and creation (annihilation) operators $c_n^{\dagger}\;(c_n)$.

The electron charge density operator reads $\hat{\rho}(\mathbf{q})= -e\sum_{\mathbf{k}nn'}\left[ \mathbf{n}\right]_{nn'}(\mathbf{k},\mathbf{k}+\mathbf{q}) c^\dagger_{n\mathbf{k}}c_{n'\mathbf{k}+\mathbf{q}}$. The corresponding matrix element  $\left[ \mathbf{n}\right]_{nn'}(\mathbf{k},\mathbf{k}+\mathbf{q})=\langle n\mathbf{k}| e^{-i\mathbf{q}\mathbf{r}}|n'\mathbf{k+\mathbf{q}}\rangle$ reduces to $\left[ \mathbf{n}\right]_{nn'}(\mathbf{k},\mathbf{k}+\mathbf{q})\approx \delta_{nn'}$ in the long-wavelength limit $|\mathbf{q}|\to 0$. Usually, it is approximated by $\delta_{nn'}$ even for finite momenta $\mathbf{q}$ \cite{nag2020,knupfer2022,husain2023,bejas2024,schultz2025,devries2025,shen2026}. We adopt the same approximation throughout this work. 

Similarly, the current operator in the $\alpha$ direction is equal to $\hat{j}_\alpha(\mathbf{q})=-e \sum_{nn'\mathbf{k}}[\mathbf{v}_\alpha]_{nn'}(\mathbf{k}+\frac{\mathbf{q}}{2})c^\dagger_{n\mathbf{k}}c_{n'\mathbf{k}+\mathbf{q}}$ \cite{mahanbook}, with the velocities obtained as $\left[ \mathbf{v}_\alpha\right] _{nn'}(\mathbf{k})=\frac{1}{\hbar}\frac{\partial \left[ \mathbf{H}_{\mathrm 0}\right] _{nn'}(\mathbf{k})}{\partial k_\alpha}$. Note that the momentum independent spin-orbit coupling (SOC) does not contribute to $\mathbf{v}$. Since the dipole operator can be expressed in terms of the charge density operator, $\hat{P}_\alpha(\mathbf{q}) = \frac{i}{q_\alpha}\hat{\rho}(\mathbf{q})$, the retarded current-dipole correlation function reads

\begin{equation}
	\begin{split}
		\chi_{\alpha \beta}(\mathbf{q},t-t')&= -\frac{i}{\hbar}\Theta(t-t')\left\langle  \left[ \hat{j}_\alpha(\mathbf{q},t),\hat{P}_\beta(-\mathbf{q},t')\right] \right\rangle =\frac{i}{q_\beta}\left\lbrace -\frac{i}{\hbar}\Theta(t-t')\left\langle  \left[ \hat{j}_\alpha(\mathbf{q},t),\hat{\rho}(-\mathbf{q},t')\right] \right\rangle\right\rbrace  \\ &=\frac{i}{q_\beta}\left\lbrace -\frac{i}{\hbar}\Theta(t-t')\sum_{nn'n''}\sum_{\mathbf{k}\mathbf{k}'}\left[\mathbf{v}_\alpha \right]_{nn''} \left( \mathbf{k}+\frac{\mathbf{q}}{2}\right) \left\langle\left[ c^\dagger_{n\mathbf{k}}c_{n''\mathbf{k}+\mathbf{q}}(t),c^\dagger_{n'\mathbf{k}'+\mathbf{q}}c_{n'\mathbf{k}'}(t')\right]\right\rangle\right\rbrace = \frac{i}{q_\beta}\chi_{\alpha 0}(\mathbf{q},t-t') \;,
			\end{split}
\end{equation}
where  $\chi_{\alpha 0}$ is the current-density correlation function, and $\left\langle \right\rangle$ denotes the ground state or thermal average. For now, we set $\frac{e^2}{V}\equiv 1$, where $V$ denotes the primitive cell volume; it will be reintroduced later through the Coulomb interaction. The sums over momenta implicitly include normalization with respect to the number of $k$-points, i.e., $\sum_{\mathbf{k}}\equiv \frac{1}{N_k}\sum_{\mathbf{k}}$.

After inclusion of the SOC, the non-interacting part of the Hamiltonian $\hat{\mathbf{H}}_{\text{0}}+\hat{\mathbf{H}}_{\text{SOC}}$ can be diagonalized and yields six bands $\left\lbrace l \right\rbrace $ with dispersions $\varepsilon_{l}\left(\mathbf{k} \right) $. The two basis sets $\left\lbrace n \right\rbrace $ and $\left\lbrace l \right\rbrace $ are related through a unitary transformation $d^\dagger_{l\mathbf{k}}=\sum_{n}^{}\left[ \mathbf{U}\right]_{nl}(\mathbf{k}) c^\dagger_{n\mathbf{k}}$. The current-density correlation function can now be expressed in terms of creation (annihilation) operators $d^\dagger_{l}\;(d_l)$ corresponding to the eigenstates of the non-interacting part of the Hamiltonian

\begin{equation} \label{eq:current-density-general}
	\begin{split} \chi_{\alpha 0}(\mathbf{q},t-t')&=-\frac{i}{\hbar}\Theta(t-t')\sum_{\mathbf{k}\mathbf{k}'}\sum_{nn'n''}\left[ \mathbf{v}_\alpha\right] _{nn''}\left( \mathbf{k}+\frac{\mathbf{q}}{2}\right) \sum_{ll'l''l'''}^{}\left[ \mathbf{U}^*\right] _{n l}(\mathbf{k})\left[ \mathbf{U}\right] _{n'' l'}(\mathbf{k}+\mathbf{q})\left[ \mathbf{U}^*\right] _{n' l''}(\mathbf{k}'+\mathbf{q})\left[ \mathbf{U}\right] _{n' l'''}(\mathbf{k}')\\
		&\times \left\langle \left[ d^\dagger_{l\mathbf{k}}d_{l'\mathbf{k}+\mathbf{q}}(t),d^\dagger_{l''\mathbf{k}'+\mathbf{q}}d_{l'''\mathbf{k}'}(t')\right] \right\rangle \;.
	\end{split}
\end{equation}
Without interactions, the above expectation value can be readily evaluated yielding the Lindhard form \cite{mihaila2011}. After the Fourier transform to the frequency/energy domain, it reads

\begin{equation}
	\begin{split} \chi^0_{\alpha 0}(\mathbf{q},\omega)&=\sum_{\mathbf{k}}\sum_{nn'n''}\left[ \mathbf{v}_\alpha\right] _{nn''}\left( \mathbf{k}+\frac{\mathbf{q}}{2}\right) \sum_{ll'}^{}\left[ \mathbf{U}^*\right] _{n l}(\mathbf{k})\left[ \mathbf{U}\right] _{n'' l'}(\mathbf{k}+\mathbf{q})\left[ \mathbf{U}^*\right] _{n' l'}(\mathbf{k}+\mathbf{q})\left[ \mathbf{U}\right] _{n' l}(\mathbf{k})\\
		&\times \frac{ f\left( \varepsilon_{l'}(\mathbf{k}+\mathbf{q})\right) - f\left( \varepsilon_l(\mathbf{k})\right) }{\omega -\left( \varepsilon_l(\mathbf{k})-\varepsilon_{l'}(\mathbf{k}+\mathbf{q})\right) + i0^+}\;.
	\end{split}
\end{equation}
Here, $f\left( \varepsilon_l(\mathbf{k})\right) =\frac{1}{e^{\beta(\varepsilon_l(\mathbf{k})-\mu)}+1}$ is the Fermi-Dirac distribution, with $\beta=(k_BT)^{-1}$ the inverse temperature and $\mu$ the chemical potential. Note that $\omega$ has units of energy.

With interactions included, the evaluation of Eq.~\eqref{eq:current-density-general} is no longer trivial. In principle, in addition to the self-energy $\boldsymbol{\Sigma}$ contributions, there are vertex corrections. Within dynamical mean-field theory (DMFT), these corrections vanish in the single-band case in the long-wavelength limit due to the current vertices being odd in momentum \cite{khurana1990}.  For the present multiband case, an exact cancellation of vertex corrections is not generally expected. Nevertheless, we anticipate that vertex corrections to the current–dipole correlation function are strongly suppressed, and in the following we thus retain only the bubble term. In Sec.~\ref{eq:cont_eq_sum_rule}, we demonstrate that the current-dipole correlation function, evaluated at the level of the bubble contribution, conserves the optical sum rule to a great extent, corroborating this expectation.

Second, even within the DMFT bubble contribution, there might exist 
processes involving all four different states $l,l',l''$, and $l'''$, the latter being described with the interacting retarded (advanced) Green's function $\mathbf{G}^R=\left[\left( \omega+\mu\right)\cdot\mathbf{I} -\mathbf{H}_{0}(\mathbf{k})-\mathbf{H}_{\rm SOC}-\boldsymbol{\Sigma}(\omega) \right] ^{-1}\;(\mathbf{G}^A = \left[ \mathbf{G}^R\right] ^*)$ and spectral function $\mathbf{A} = \pi^{-1}\text{Im}\mathbf{G}^R$. The correlation function can then be evaluated as \cite{worm2021}

\begin{equation} \label{eq:current-density-bubble}
	\begin{split}
		\chi_{\alpha0}(\mathbf{q},\omega)&=\sum_{\mathbf{k}}\sum_{nn'n''} \left[ \mathbf{v}_\alpha\right] _{nn''}\left( \mathbf{k}+\frac{\mathbf{q}}{2}\right)  \sum_{ll'l''l'''}^{}\left[ \mathbf{U}^*\right] _{n l}(\mathbf{k})\left[ \mathbf{U}\right] _{n'' l'}(\mathbf{k}+\mathbf{q})\left[ \mathbf{U}^*\right] _{n' l''}(\mathbf{k}+\mathbf{q})\left[ \mathbf{U}\right] _{n' l'''}(\mathbf{k})\\
		&\times\int_{-\infty}^{+\infty}d\varepsilon f(\varepsilon)\left\lbrace \left[ \mathbf{A}\right] _{l'''l}(\mathbf{k},\varepsilon)\left[ \mathbf{G}^R\right] _{l'l''}(\mathbf{k}+\mathbf{q},\varepsilon+\omega)+\left[ \mathbf{A}\right] _{l'l''}(\mathbf{k}+\mathbf{q},\varepsilon)\left[ \mathbf{G}^A\right] _{l'''l}(\mathbf{k},\varepsilon-\omega)\right\rbrace \\ 
        &=\sum_{\mathbf{k}}\sum_{nn'n''} \left[ \mathbf{v}_\alpha\right] _{nn''}\left( \mathbf{k}+\frac{\mathbf{q}}{2}\right) 
        \\
		&\times\int_{-\infty}^{+\infty}d\varepsilon f(\varepsilon)\left\lbrace \left[ \mathbf{G}^R\right] _{n''n'}(\mathbf{k}+\mathbf{q},\varepsilon+\omega)\left[ \mathbf{A}\right] _{n'n}(\mathbf{k},\varepsilon)+\left[ \mathbf{A}\right] _{n''n'}(\mathbf{k}+\mathbf{q},\varepsilon)\left[ \mathbf{G}^A\right] _{n'n}(\mathbf{k},\varepsilon-\omega)\right\rbrace  \;,
	\end{split}
\end{equation}
where $f(\varepsilon)=\frac{1}{e^{\beta\varepsilon}+1}$.

\subsection{Fourier transforms and numerical evaluation}

One can efficiently evaluate Eq.~\eqref{eq:current-density-bubble} using Fourier transforms. In particular, let $\chi^{(1)}_{nn'n''}(\mathbf{q},\omega)$ represent the first contribution in the second line in Eq.~\eqref{eq:current-density-bubble}. The corresponding Fourier transform reads

\begin{equation} \label{eq:time_transform}
	\begin{split}
		&\chi^{(1)}_{nn'n''}(\mathbf{q},t)=\int_{-\infty}^{+\infty}d\omega e^{-i\omega t}\chi^{(1)}_{nn'n''}(\mathbf{q},\omega)=\ \int_{-\infty}^{+\infty}d\omega e^{-i\omega t}\int_{-\infty}^{+\infty}d\varepsilon f(\varepsilon) \left[ \mathbf{G}^R\right] _{n''n'}(\mathbf{k}+\mathbf{q},\varepsilon+\omega)\left[ \mathbf{A}\right] _{n'n}(\mathbf{k},\varepsilon)\\
		&=\int_{-\infty}^{+\infty}d\omega e^{-i\omega t} \left[ \mathbf{G}^R\right] _{n''n'}(\mathbf{k}+\mathbf{q},\omega) \int_{-\infty}^{+\infty}d\varepsilon e^{i\varepsilon t}f(\varepsilon)\left[ \mathbf{A}\right] _{n'n}(\mathbf{k},\varepsilon)\\
		&
		= \int_{-\infty}^{+\infty}d\omega e^{-i\omega t} \left[ \mathbf{G}^R\right] _{n''n'}(\mathbf{k}+\mathbf{q},\omega) \int_{-\infty}^{+\infty}d\varepsilon e^{-i\varepsilon t}f(-\varepsilon)\left[ \mathbf{A}\right] _{n'n}(\mathbf{k},-\varepsilon)
		=\left[ \mathbf{G}^R\right] _{n''n'}(\mathbf{k}+\mathbf{q},t)\left\lbrace  f\left[ \mathbf{A}\right] _{n'n}\right\rbrace  (\mathbf{k},t)\;.
	\end{split}
\end{equation}
Note that $t$ in Eq.~\eqref{eq:time_transform} and the subsequent equations has units of inverse energy. Applying the same procedure to the second contribution, together with the first contribution one then obtains

\begin{equation}
	\begin{split}
		\chi_{\alpha0}(\mathbf{q},\omega)&=\sum_{\mathbf{k}}\sum_{nn'n''} \left[ \mathbf{v}_\alpha\right] _{nn''}\left( \mathbf{k}+\frac{\mathbf{q}}{2}\right)  \\
		&\times\int_{-\infty}^{+\infty}dt e^{i\omega t} \left\lbrace \left[ \mathbf{G}^R\right] _{n''n'}(\mathbf{k}+\mathbf{q},t)\left\lbrace  f\left[ \mathbf{A}\right] _{n'n}\right\rbrace  (\mathbf{k},t)+\left\lbrace  f\left[ \mathbf{A}\right] _{n''n'}\right\rbrace  (\mathbf{k}+\mathbf{q},t)\left[ \mathbf{G}^A\right] _{n'n}(\mathbf{k},t)\right\rbrace  \;.
	\end{split}
\end{equation}

We can exploit the same Fourier transform scheme to evaluate the sum over $\mathbf{k}$ if we take $[\mathbf{v}_\alpha]_{nn'}(\mathbf{k}+\frac{\mathbf{q}}{2})\approx\frac{1}{2}\left\lbrace \left[ \mathbf{v}_\alpha\right] _{nn'}(\mathbf{k}+\mathbf{q})+\left[ \mathbf{v}_\alpha\right] _{nn'}(\mathbf{k}) \right\rbrace$. This gives four contributions

\begin{equation}
	\begin{split}
		& \sum_{nn'n''}\left\lbrace \left[ \mathbf{v}_\alpha\right] _{nn''}(\mathbf{k})+\left[ \mathbf{v}_\alpha\right] _{nn''}(\mathbf{k}+\mathbf{q}) \right\rbrace\left\lbrace \left[ \mathbf{G}^R\right] _{n''n'}(\mathbf{k}+\mathbf{q},t)\left\lbrace  f\left[ \mathbf{A}\right] _{n'n}\right\rbrace  (\mathbf{k},t) + \left\lbrace  f\left[ \mathbf{A}\right] _{n''n'}\right\rbrace  (\mathbf{k}+\mathbf{q},t)\left[ \mathbf{G}^A\right] _{n'n}(\mathbf{k},t)\right\rbrace \\
		&= \sum_{nn'} \left\lbrace \left[ \mathbf{G}^R\right] _{nn'}(\mathbf{k}+\mathbf{q},t)\left\lbrace  f\left[ \mathbf{A}\mathbf{v}_\alpha\right] _{n'n}\right\rbrace  (\mathbf{k},t)  +\left[ \mathbf{v}_\alpha\mathbf{G}^R\right] _{nn'}(\mathbf{k}+\mathbf{q},t) \left\lbrace  f\left[ \mathbf{A}\right] _{n'n}\right\rbrace  (\mathbf{k},t) \right. \\
		&\left.+ \left\lbrace  f\left[ \mathbf{A}\right] _{nn'}\right\rbrace  (\mathbf{k}+\mathbf{q},t)\left[ \mathbf{G}^A\mathbf{v}_\alpha\right] _{n'n}(\mathbf{k},t)+ \left\lbrace  f\left[ \mathbf{v}_\alpha\mathbf{A}\right] _{nn'}\right\rbrace  (\mathbf{k}+\mathbf{q},t)\left[ \mathbf{G}^A\right] _{n'n}(\mathbf{k},t)\right\rbrace \;.
	\end{split}
\end{equation}
By looking, for example, at the first contribution, we can further write

\begin{equation}
	\begin{split}
		&\chi_{nn'}(\mathbf{r},t)=\sum_{\mathbf{q}}e^{-i\mathbf{q}\mathbf{r}}\chi_{nn'}(\mathbf{q},t)=\sum_{\mathbf{q}}e^{-i\mathbf{q}\mathbf{r}}\sum_{\mathbf{k}} \left[ \mathbf{G}^R\right] _{nn'}(\mathbf{k}+\mathbf{q},t)\left\lbrace  f\left[ \mathbf{A}\mathbf{v}_\alpha\right] _{n'n}\right\rbrace  (\mathbf{k},t)  \\
		&=  \sum_{\mathbf{q}}e^{-i\mathbf{q}\mathbf{r}}\left[ \mathbf{G}^R\right] _{nn'}(\mathbf{q},t) \sum_{\mathbf{k}} e^{i\mathbf{k}\mathbf{r}}\left\lbrace  f\left[ \mathbf{A}\mathbf{v}_\alpha\right] _{n'n}\right\rbrace  (\mathbf{k},t)  =\sum_{\mathbf{q}}e^{-i\mathbf{q}\mathbf{r}}\left[ \mathbf{G}^R\right] _{nn'}(\mathbf{q},t) \sum_{\mathbf{k}} e^{-i\mathbf{k}\mathbf{r}}\left\lbrace  f\left[ \mathbf{A}\mathbf{v}_\alpha\right] _{n'n}\right\rbrace  (-\mathbf{k},t)   \\
		&=\left[ \mathbf{G}^R\right] _{nn'}(\mathbf{r},t) \left\lbrace  f\left[ \mathbf{A}\mathbf{v}_\alpha\right] _{n'n}\right\rbrace  (\mathbf{r},t)  \;.
	\end{split}
\end{equation}
Using the same procedure for all contributions, one gets

\begin{equation}
	\begin{split}
		\chi_{\alpha0}(\mathbf{q},\omega)
		&=\frac{1}{2}\sum_{nn'}\int_{-\infty}^{+\infty}dt e^{i\omega t}\sum_{\mathbf{r}}e^{i\mathbf{q}\mathbf{r}} \left\lbrace \left[ \mathbf{G}^R\right] _{nn'}(\mathbf{r},t) \left\lbrace  f\left[ \mathbf{A}\mathbf{v}_\alpha\right] _{n'n}\right\rbrace  (\mathbf{r},t)+\left[ \mathbf{v}_\alpha\mathbf{G}^R\right] _{nn'}(\mathbf{r},t) \left\lbrace  f\left[ \mathbf{A}\right] _{n'n}\right\rbrace  (\mathbf{r},t)\right.\\
		&\left.+\left\lbrace  f\left[ \mathbf{A}\right] _{nn'}\right\rbrace  (\mathbf{r},t)\left[ \mathbf{G}^A\mathbf{v}_\alpha\right] _{n'n}(\mathbf{r},t)+ \left\lbrace  f\left[ \mathbf{v}_\alpha\mathbf{A}\right] _{nn'}\right\rbrace  (\mathbf{r},t)\left[ \mathbf{G}^A\right] _{n'n}(\mathbf{r},t)\right\rbrace\\
		&=\frac{1}{2}\text{Tr}\left[ \int_{-\infty}^{+\infty}dt e^{i\omega t}\sum_{\mathbf{r}}e^{i\mathbf{q}\mathbf{r}}   \left(  \left\lbrace  \mathbf{G}^R\left[f\mathbf{A}\mathbf{v}_\alpha\right] \right\rbrace  + \left\lbrace \left[ \mathbf{v}_\alpha\mathbf{G}^R\right]  \left[f\mathbf{A}\right] \right\rbrace + \left\lbrace \left[ f\mathbf{A}\right] \left[\mathbf{G}^A\mathbf{v}_\alpha\right] \right\rbrace + \left\lbrace\left[ f \mathbf{v}_\alpha\mathbf{A}\right] \mathbf{G}^A\right\rbrace  \right) (\mathbf{r},t)\right] \;.
	\end{split}
\end{equation}

In our calculations, we used the momentum grid of $N_k=N_{k_x}\times N_{k_y}\times N_{k_z}=128\times128\times1$ points. The energy grid had $N_\omega=7501$ points in the range $[-15,15]$ eV. Before performing the Fourier transforms, the data was zero padded so that the total number of energy points became $2^{14}$. In addition to the DMFT self-energy, an additional tiny broadening of $10^{-7}$ eV was added by hand.

\section{Continuity equation and optical sum rule} \label{eq:cont_eq_sum_rule}

In order to study plasmon excitations, we need to calculate the density-density correlation function $\chi_{00}$. By exploiting the continuity equation, it can be obtained from the conductivity tensor\,\cite{schrieffer2018}

\begin{equation} \label{eq:continuity_equation}
	\chi_{00}(\mathbf{q},\omega) = \frac{\hbar}{i\omega}\sum_{\alpha\beta}q_\alpha\sigma_{\alpha\beta}(\mathbf{q},\omega)q_\beta\;,
\end{equation}
which in turn can be calculated either from the current-dipole $ \chi_{\alpha\beta}$, current-density  $\chi_{\alpha0}$, or current-current correlation function  $\tilde{\chi}_{\alpha\beta}$

\begin{equation} \label{eq:sigma_vs_correlation}
	\begin{split}
			\sigma_{\alpha\beta}(\mathbf{q},\omega)& = \chi_{\alpha\beta}(\mathbf{q},\omega)=\frac{i}{q_\beta}\chi_{\alpha0}(\mathbf{q},\omega) =\frac{i\hbar}{\omega}\left( \tilde{\chi}_{\alpha\beta}(\mathbf{q},\omega) + \frac{e^2n_{\alpha\beta}(\mathbf{q})}{m}\right)\;. 
	\end{split}
\end{equation}
Note that the current-current correlation function gives just the paramagnetic contribution to the conductivity, while the diamagnetic term $\frac{e^2n_{\alpha\beta}(\mathbf{q})}{m}$  needs to be additionally evaluated. Hence, we focus on the current-dipole/current-density correlation function, and by combining Eqs.~\eqref{eq:continuity_equation} and~\eqref{eq:sigma_vs_correlation} we can write

\begin{equation} \label{eq:chi_dens_vs_chi_jrhp}
		\chi_{00}(\mathbf{q},\omega) = \frac{\hbar}{\omega}\sum_{\alpha}q_\alpha\chi_{\alpha0}(\mathbf{q},\omega)\;.
\end{equation}

 \begin{figure}
	\includegraphics[width=0.75\textwidth]{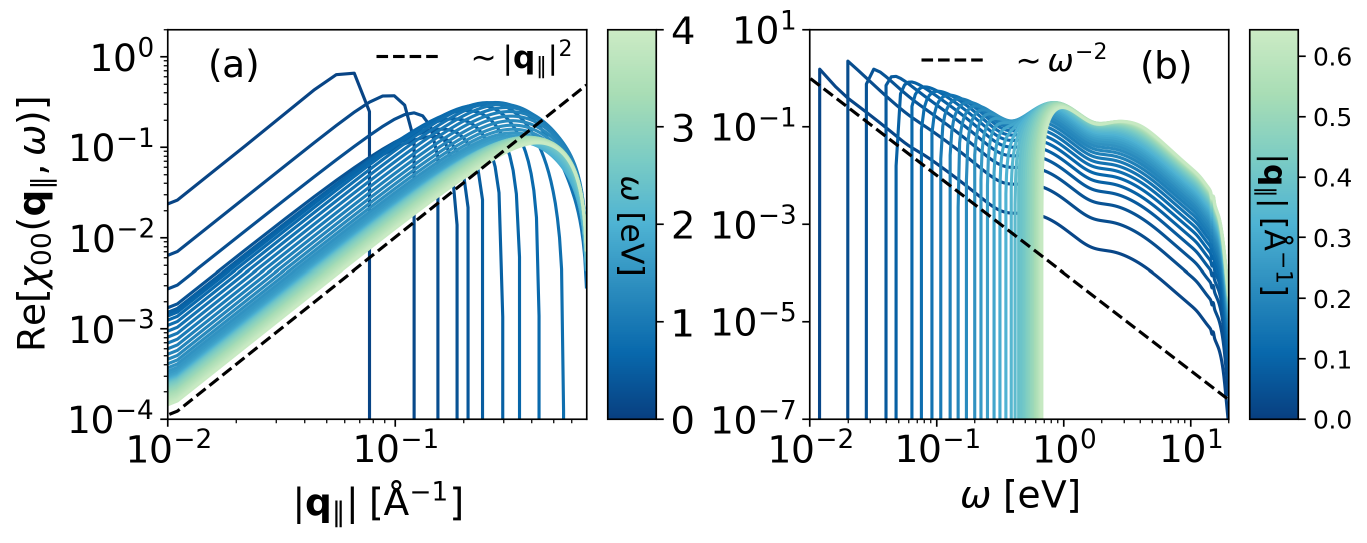}
	\renewcommand{\figurename}{Fig.}
	\caption{Real part of the DFT+DMFT density-density correlation function $\chi_{00}(|\mathbf{q}_\Vert|,\omega)$ as a function of (a) $|\mathbf{q}_\Vert|$ at fixed $\omega$ and (b) $\omega$ at fixed fixed $|\mathbf{q}_\Vert|$. The lines $|\mathbf{q}_\Vert|^2$ and $\omega^{-2}$  are shown in (a) and (b), respectively, as guides to the eye to demonstrate the $|\mathbf{q}_\Vert|^2/\omega^2$ scaling of Re$\chi_{00}(|\mathbf{q}_\Vert|,\omega)$ in the long-wavelength limit.}
	\label{fig:q_w_dep_density}
\end{figure}

In Figs.~\ref{fig:q_w_dep_density}(a) and (b), we show the in-plane momentum $|\mathbf{q}_\Vert|$ dependence for a fixed $\omega$, and the  $\omega$ dependence for a fixed $|\mathbf{q}_\Vert|$ of the real part of the DFT+DMFT density-density correlation function $\chi_{00}(|\mathbf{q}_\Vert|,\omega)$ calculated by means of Eq.~\eqref{eq:chi_dens_vs_chi_jrhp}. Our results clearly exhibit the $|\mathbf{q}_\Vert|^2/\omega^2$ scaling in the long-wavelength limit. Moreover, it follows from Eq.~\eqref{eq:chi_dens_vs_chi_jrhp} that $\lim_{|\mathbf{q}_\Vert|\to0}\chi_{00}(|\mathbf{q}_\Vert|,\omega)\to 0$ for $\omega\geq0$, in accordance with the charge conservation law.

In general, to ensure charge conservation, the self-energy corrections should be accompanied with the corresponding two-particle vertex contributions generated from  the same Luttinger-Ward functional when calculating correlation functions\,\cite{schrieffer2018}. Indeed, when $\chi_{00}(|\mathbf{q}_\Vert|,\omega) $ is calculated directly by considering just the bubble contribution with the DMFT self-energy, the plasmon energy diverges in the long-wavelength limit. Instead, when the current-dipole correlation function is calculated first, and the continuity equation is exploited to obtain the density response, a finite plasmon energy is restored, even without two-particle vertex contributions.

To further quantify the quality of our approach, we evaluate $\frac{2\hbar}{\pi\epsilon_0}\int_{0}^{\infty}d\omega\text{Re}\sigma(\omega)\equiv (\hbar\Omega_p)^2$. Here, $\epsilon_0$ is the vacuum permittivity. In a conserving approximation, this integral satisfies the restricted optical sum rule. It is determined by the kinetic energy, or more precisely by $(\hbar\Omega_{SR})^2=\frac{e^2}{\epsilon_0V}\lim_{|\mathbf{q}|\to 0}\sum_{\alpha\beta}\frac{q_\alpha q_\beta}{|\mathbf{q}|^2}\left\langle\tau_{\alpha\beta}\right\rangle$, where $\left\langle\tau_{\alpha\beta}\right\rangle=\sum_{nn'\mathbf{k}}\frac{\partial^2 \left[ \mathbf{H}_{\mathrm 0}\right] _{nn'}(\mathbf{k})}{\partial k_\alpha\partial k_\beta}\int d\varepsilon f(\varepsilon-\mu)\left[ \mathbf{A}\right] _{nn'}(\mathbf{k},\varepsilon)$.
We obtain $(\hbar\Omega_{p})^2=(3.52\;\text{eV})^2\approx 0.9\; (\hbar\Omega_{SR})^2$. Evidently, our approach captures 
90\% of the restricted optical sum rule even in the absence of vertex corrections. This level of agreement is highly satisfactory, particularly given that a direct evaluation of the density–density correlation function without vertex corrections leads to a complete violation of the sum rule.

\section{Memory function and plasmon energy}

\begin{figure}
	\includegraphics[width=0.75\textwidth]{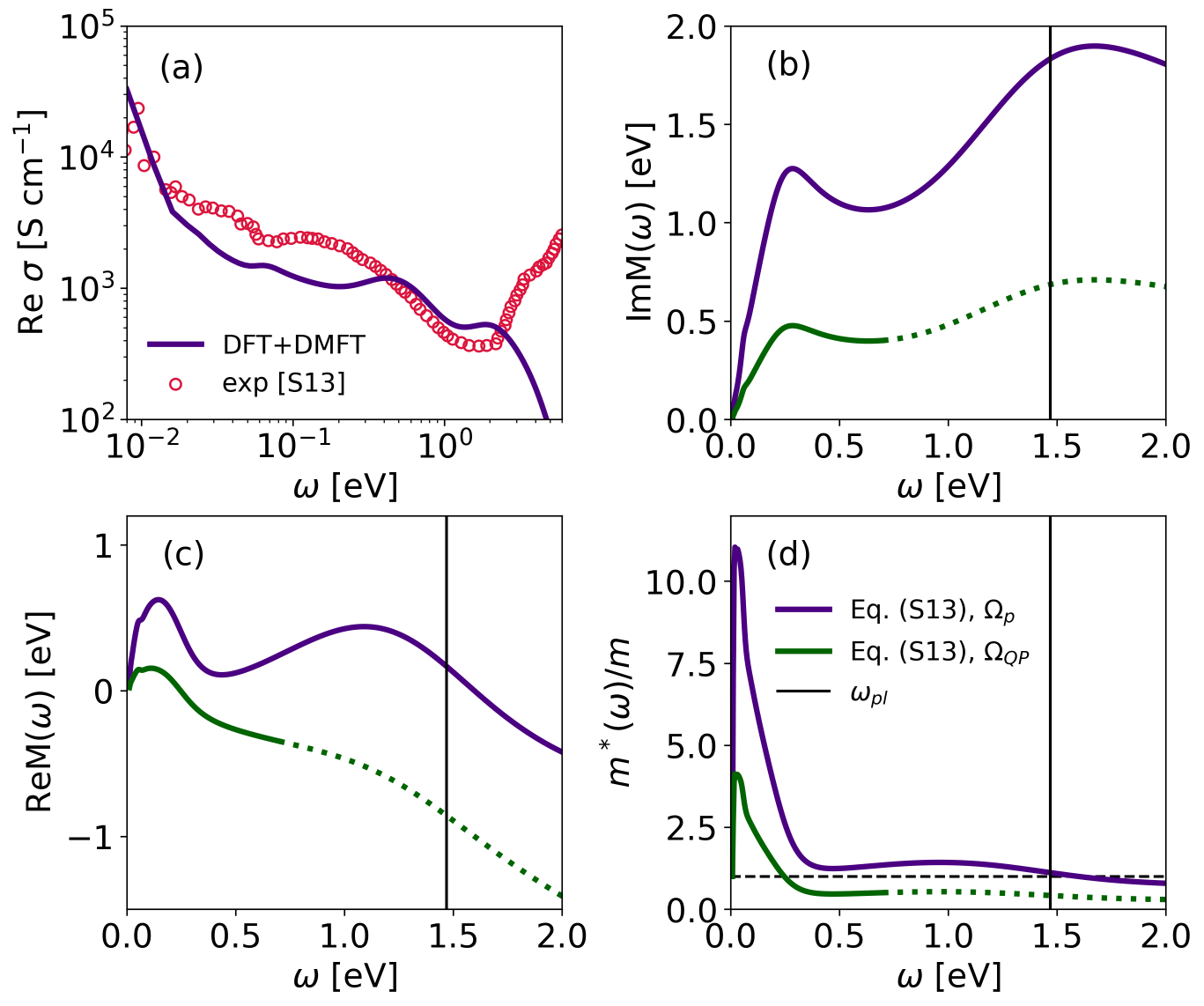}
	\renewcommand{\figurename}{Fig.}
	\caption{(a) DFT+DMFT (purple line) optical conductivity compared to experiment (red circles) from Ref.~\cite{stricker2014}. (b) Imaginary and (c) real part of the DFT+DMFT memory function $M(\omega)$ given by Eq.~\eqref{eq:sigma_memory_function} together with (d) the optical effective mass $m^*(\omega)/m=1+\frac{\text{Re}M(\omega)}{\omega}$.  The purple lines correspond to the case $\Omega=\Omega_p$, while the green lines to the case $\Omega=\Omega_{QP}$. The optical plasmon energy is shown for reference as a black vertical line.}
	\label{fig:memory_function}
\end{figure}

Using the DFT+DMFT-calculated charge response, we can evaluate the dielectric function $\varepsilon(\mathbf{q},\omega)=\varepsilon_\infty - V({\mathbf{q}})\chi_{00}(\mathbf{q,\omega})$, and in turn the optical conductivity $\sigma(\omega) = i\frac{\epsilon_0\omega}{\hbar}\left[ \varepsilon_\infty-\varepsilon(0,\omega)\right]$. For a layered system, the Coulomb interaction is given by $V(\mathbf{q}_\Vert,q_\bot)=\frac{e^2}{2\epsilon_0 |\mathbf{q}_{\Vert}|A}\frac{\sinh\left( |\mathbf{q}_{\Vert}|d \right) }{\cosh( |\mathbf{q}_{\Vert}|)-\cos(q_\bot d)}$. We take an experimental value of $\varepsilon_\infty\approx2.3$ \cite{stricker2014}. In Fig.~\ref{fig:memory_function}(a), we depict our DFT+DMFT results for the optical conductivity, which shows a good agreement with the optical measurements\,\cite{stricker2014}.

 A convenient way to address the influence of correlation effects on optical conductivity is through the memory function $M(\omega)$ \cite{gotze1972}. It is defined in terms of the optical conductivity as
\begin{equation} \label{eq:sigma_memory_function}
	\sigma(\omega) = \frac{i\epsilon_0\hbar\Omega^2}{\omega+M(\omega)}\equiv\frac{i\epsilon_0\hbar\Omega^2}{\frac{m^*(\omega)}{m}\omega+i\Gamma(\omega)}\;,
\end{equation}
 where the optical effective mass is given by the real part of the memory function $m^*(\omega)/m=1+\frac{\text{Re}M(\omega)}{\omega}$, while the imaginary part determines the scattering rate $\Gamma(\omega)=\text{Im}M(\omega)$.
 
The memory function is commonly employed within the extended Drude model to address the energy dependence of the scattering rate and effective mass of itinerant charge carriers. In this context, $\Omega$ is associated with the intraband spectral weight. The situation in Sr$_2$RuO$_4$, however, is more intricate. First, due to strong correlations, the single-particle spectral weight is redistributed between the quasiparticle bands and incoherent features, with only the former contributing to the spectral weight of the itinerant carriers. Second,  the presence of multiple bands near the Fermi level gives rise to low-energy interband transitions that also contribute to the optical conductivity. As a result, both  $\Omega$ and memory function should be regarded as effective quantities describing the total low-energy response. In particular, from Fig.~1(b) of the main text, we can estimate the upper bound to the largest energy transition within the renormalized quasiparticle bands to be $\omega_{upper}\approx 0.7$ eV. Therefore, in the spirit of the extended Drude model we may define $(\hbar\Omega_{QP})^2\equiv \frac{2\hbar}{\pi\varepsilon_0}\int_{0}^{\omega_{upper}}d\omega\text{Re}\sigma(\omega) \approx 0.37\;(\hbar\Omega_p)^2$. The corresponding memory function and optical effective mass, obtained by using $\Omega=\Omega_{QP}$ in Eq.~\eqref{eq:sigma_memory_function}, are shown in Fig.~\ref{fig:memory_function} with the green lines. For $\omega> \omega_{upper}$, these curves are plotted as dotted lines to emphasize that they represent effective quantities constructed to capture the low-energy response in the range $\omega < \omega_{upper}$.

To capture the high-energy plasmon response at $\omega_{pl}\approx 1.47$ eV [indicated by the black vertical line in Fig.~\ref{fig:memory_function}], we introduce an alternative effective memory function by using the total spectral weight in Eq.~\eqref{eq:sigma_memory_function}, i.e., setting $\Omega=\Omega_p$. The resulting memory function and optical effective mass are shown in Fig.~\ref{fig:memory_function} with the purple lines. We note that a qualitatively similar structure of the memory function has been observed in other correlated materials \cite{hwang2004,Heumen2009,conte2012}. There, the first peak in the real part of $M(\omega)$ is usually associated with a bosonic (e.g., magnetic) mode, while a broad continuum at higher energies of electronic origin. Here, we show that such a structure naturally occurs in the presence of strong local electron correlations, without invoking any  bosonic mode.

Fig.~\ref{fig:memory_function}(d) indicates that the optical effective mass is close to unity near the optical plasmon energy. In contrast, it is significantly enhanced at low energies, relevant for the lower-energy, acoustic-like layered plasmons at finite $q_\bot$.
Moreover, Fig.~\ref{fig:memory_function}(b) shows that the imaginary part of the memory function is comparable in magnitude to the plasmon resonance itself. This translates to the substantial plasmon broadening, cf.\ \cite{Si2021}, but also affects the actual plasmon energy. More explicitly, using Eq.~\eqref{eq:sigma_memory_function}, the dielectric function can be written in terms of the memory function that captures the full response

\begin{equation} \label{eq:epsilon_memory_function}
	\varepsilon(\omega)=\varepsilon_\infty - \frac{(\hbar\Omega_p)^2}{\omega(\omega+M(\omega))}=\varepsilon_\infty - \frac{(\hbar\Omega_p)^2}{\frac{m^*(\omega)}{m}\omega^2+i\omega\Gamma(\omega)}\;.
\end{equation}
The plasmon energy can now be obtained as the zero crossing of the real part of the dielectric function, $\text{Re}\;\varepsilon(\omega_{pl})=0$, giving

\begin{equation} \label{eq:plasmon_pole_solution}
     \omega_{pl}=\sqrt{\frac{(\hbar\Omega_p)^2}{\varepsilon_\infty m^*(\omega_{pl})/m}-\left(\frac{\Gamma(\omega_{pl})}{m^*(\omega_{pl})/m}\right)^2}\;.
\end{equation}
Using $\varepsilon_\infty=2.3$, together with $m^*(\omega_{pl})/m\approx 1.11$ and $\Gamma(\omega_{pl})\approx 1.83$ eV extracted from Fig.~\ref{fig:memory_function}, and $\hbar\Omega_p \approx 3.52$ eV, we recover $\omega_{pl}\approx 1.47$ eV. 
Note that, irrespective of the choice of $\Omega$ in Eq.~\eqref{eq:sigma_memory_function} used to define an effective memory function, the ratios $\Omega^2/[m^*(\omega)/m]$ and $\Gamma(\omega)/[m^*(\omega)/m]$ remain invariant, provided the same $\sigma(\omega)$ is used.

\begin{figure}
	\includegraphics[width=1\textwidth]{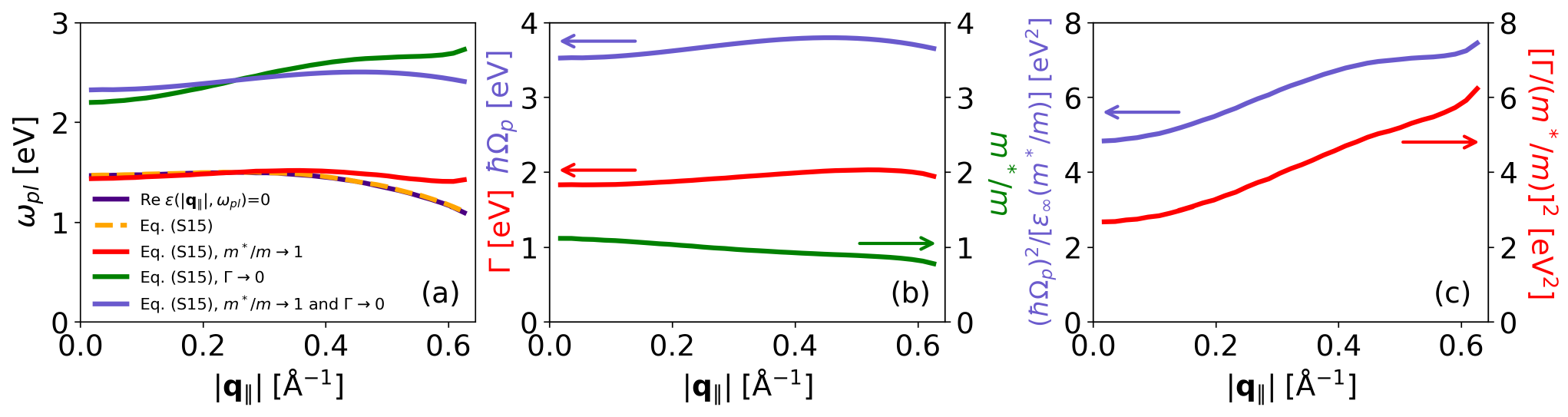}
	\renewcommand{\figurename}{Fig.}
	\caption{(a) Plasmon dispersion obtained from the condition $\text{Re}\varepsilon(|\mathbf{q}_\Vert|,\omega_{pl}(|\mathbf{q}_\Vert|))=0$ (solid purple line) and from the momentum-dependent analogue of Eq.~\eqref{eq:plasmon_pole_solution} (dashed orange line). For comparison, results are also shown for the cases $m^*(|\mathbf{q}_\Vert|,\omega_{pl}(|\mathbf{q}_\Vert|))/m\to 1$ (solid red line), $\Gamma(|\mathbf{q}_\Vert|,\omega_{pl}(|\mathbf{q}_\Vert|))\to 0$ (solid green line), and $[m^*(|\mathbf{q}_\Vert|,\omega_{pl}(|\mathbf{q}_\Vert|))/m\to 1;\Gamma(|\mathbf{q}_\Vert|,\omega_{pl}(|\mathbf{q}_\Vert|))\to 0]$ (solid blue line). (b) Momentum dependence of $\hbar\Omega_p(\mathbf{q})$ (solid blue line), $\Gamma(|\mathbf{q}_\Vert|,\omega_{pl}(|\mathbf{q}_\Vert|))$ (solid red line), and $m^*(|\mathbf{q}_\Vert|,\omega_{pl}(|\mathbf{q}_\Vert|))/m$ (solid green line). (c) Momentum dependence of $(\hbar\Omega_p)^2/[\varepsilon_\infty(m^*/m)](|\mathbf{q}_\Vert|,\omega_{pl}(|\mathbf{q}_\Vert|))$ (solid blue line) and $[\Gamma/(m^*/m)]^2(|\mathbf{q}_\Vert|,\omega_{pl}(|\mathbf{q}_\Vert|))$ (solid red line).}
	\label{fig:finite_q_analysis}
\end{figure}

We can further generalize Eq.~\eqref{eq:sigma_memory_function} to finite $\mathbf{q}$, giving a momentum-dependent optical effective mass $m^*(\mathbf{q},\omega)/m=1+\frac{\text{Re}M(\mathbf{q},\omega)}{\omega}$ and a scattering rate $\Gamma(\mathbf{q},\omega)=\text{Im}M(\mathbf{q},\omega)$, where $M(\mathbf{q},\omega) = \frac{i\epsilon_0\hbar\Omega_p(\mathbf{q})^2}{\sigma(\mathbf{q},\omega)}-\omega$ and $(\hbar\Omega_p(\mathbf{q}))^2=\frac{2\hbar}{\pi\epsilon_0}\int_{0}^{\infty}d\omega\text{Re}\sigma(\mathbf{q},\omega)$. In the following, we set $q_\bot=0$ and retain only the $|\mathbf{q}_\Vert|$ dependence. Using the plasmon dispersion obtained from the condition $\text{Re}\varepsilon(|\mathbf{q}_\Vert|,\omega_{pl}(|\mathbf{q}_\Vert|))=0$ [solid purple line in Fig.~\ref{fig:finite_q_analysis}(a)], we can extract $m^*(|\mathbf{q}_\Vert|,\omega_{pl}(|\mathbf{q}_\Vert|))/m$ and $\Gamma(|\mathbf{q}_\Vert|,\omega_{pl}(|\mathbf{q}_\Vert|))$, shown in Fig.~\ref{fig:finite_q_analysis}(b) together with $\hbar\Omega_p(|\mathbf{q}_\Vert|)$. The momentum-dependent analogue of Eq.~\eqref{eq:plasmon_pole_solution} now reproduces the full plasmon dispersion [dashed orange line in Fig.~\ref{fig:finite_q_analysis}(a)] and reveals that, although damping is crucial for obtaining the correct plasmon energy, sizable negative dispersion at large $|\mathbf{q}_\Vert|$ emerges only when all four quantities$-\Omega_p$, $\varepsilon_\infty$, $m^*/m$, and $\Gamma-$are treated on an equal footing. This highlights that a consistent determination and interpretation of the plasmon energy requires knowledge of all four quantities$-\Omega_p$, $\varepsilon_\infty$, $m^*/m$, and $\Gamma$; none of them can be treated as a free parameter. 

\begin{figure}
	\includegraphics[width=0.75\textwidth]{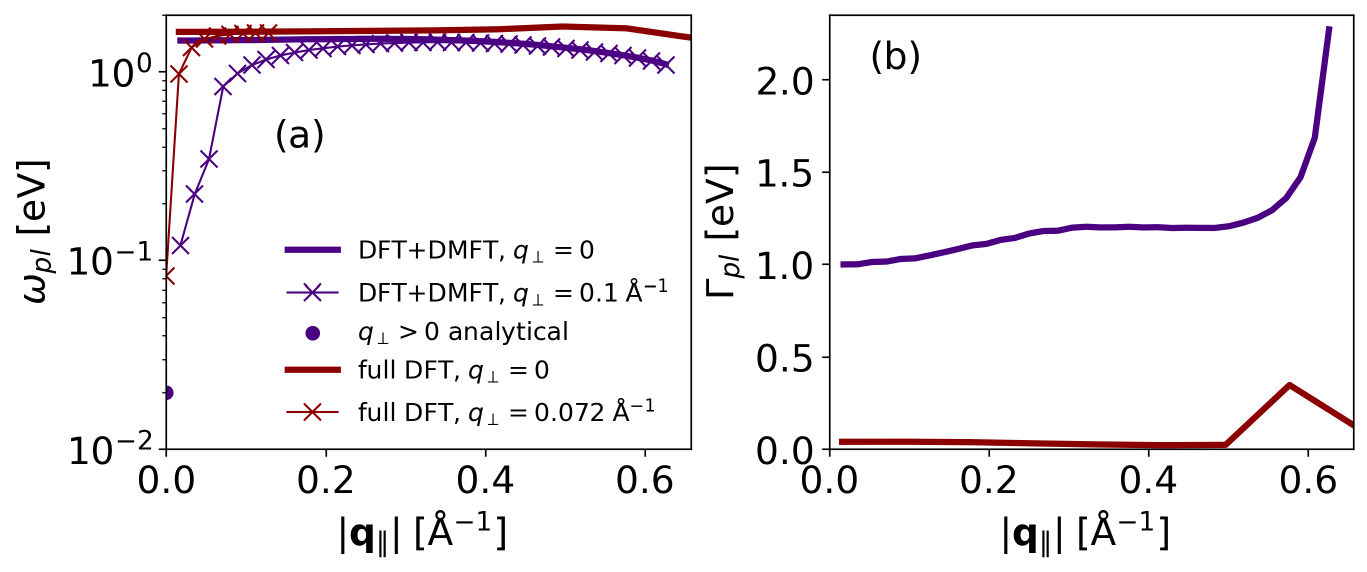}
	\renewcommand{\figurename}{Fig.}
	\caption{(a) DFT+DMFT plasmon dispersion for $q_\bot=0$ (solid purple line) and $q_\bot=0.1\;\text{\AA}^{-1}$  (purple line with x markers), together with the analytical estimate of the DFT+DMFT plasmon energy for  $|\mathbf{q}_\Vert|=0$ and $q_\bot>0$, $\omega_{pl}(|\mathbf{q}_\Vert|=0, q_\bot>0)\sim 20$ meV, and full DFT plasmon dispersion for $q_\bot=0$ (solid red line) and $q_\bot=0.072\;\text{\AA}^{-1}$  (red line with x markers). (b) DFT+DMFT (purple line) vs full DFT (red line) plasmon width.}
	\label{fig:plasmon_dft_vs_dmft}
\end{figure}

In addition, in Fig.~\ref{fig:plasmon_dft_vs_dmft}, we compare the DFT+DMFT results with the \textit{full} DFT results, where the latter involves electron bands beyond the three-band model used in the DFT+DMFT calculation$-$hence we coin it full DFT$-$but does not include the DMFT self-energy. The corresponding ground state full DFT calculations, needed for the plasmon calculations, were performed with the GPAW package~\cite{gpaw}. The plane-wave basis set with the energy cutoff of 45\,Ry and Perdew–Burke–Ernzerhof~(PBE) exchange-correlation functional were used. In order to relax the interlayer distances in Sr$_2$RuO$_4$ we have additionally used dispersion corrections with the vdW-DF functional. The unit cell parameter and atomic positions were relaxed until the forces were below $10^{-6}$\,Ry/$a_0$, with $a_0$ being the Bohr radius. 
The in-plane and out-of-plane unit cell sizes were $a=3.922$\,\AA~ and $c=13.03$\,\AA.
The momentum grids for ground-state density calculations were $\mathbf{k}=6\times 6\times 2$. The denser grids needed for the corresponding dielectric function calculations~\cite{yan2011} were $\mathbf{k}=120\times 120\times 20$, where we used a small phenomenological broadening of $\eta=10$\,meV.

Evidently, only with the DMFT correlations the plasmon width reaches 1 eV (purple line in Fig.~\ref{fig:plasmon_dft_vs_dmft}(b)) as measured in experiments. On the other hand, although the damping $\Gamma$ is negligibly small in the full DFT calculation and correlation-induced renormalization of the optical effective mass is absent, the resulting optical plasmon energy is in reasonable agreement with that obtained when correlations are included. The situation changes significantly for the layered plasmon dispersion at finite $q_\bot>0$, where the plasmon energy in the DFT+DMFT calculation is nearly an order of magnitude smaller than in the full DFT case in the long-wavelength limit. In this low-energy regime, the enhancement of the optical effective mass [Fig.~\ref{fig:memory_function}(d)] likely plays a big role in giving rise to this discrepancy. However, as indicated by Eq.~\eqref{eq:plasmon_pole_solution} and demonstrated in Fig.~\ref{fig:finite_q_analysis}, all four quantities$-\Omega_p$, $\varepsilon_\infty$, $m^*/m$, and $\Gamma-$contribute to determining the actual plasmon energy.

While the absolute plasmon energy is governed by the interplay of the aforementioned parameters, the momentum dependence of the plasmon dispersion in the long-wavelength limit follows from charge conservation. As discussed in Sec.~\ref{eq:cont_eq_sum_rule}, in the long-wavelength limit $\text{Re}\chi_{00}(\mathbf{q},\omega)\sim\frac{|\mathbf{q}|^2}{\omega^2}$,  while the Coulomb interaction in a layered system approaches a constant for $q_\bot>0$.  The resulting plasmon therefore exhibits a linear dispersion $\omega_{pl}(|\mathbf{q}_\Vert|\to0, q_\bot>0)\approx \hbar v_{pl}(q_\bot)|\mathbf{q}_\Vert|$, as can be seen in Fig.~4(a) of the main text.

A complete description of the low-energy plasmon dispersion in the limit $ |\mathbf{q}_\Vert|\to0$ and $q_\bot>0$, however, requires a reassessment of the above discussion with the inclusion of electron hopping along the out-of-plane direction. Although small compared to the in-plane delocalization energy \cite{beck2025}, this hopping becomes comparable to 
$\omega_{pl}(\mathbf{q}_\Vert, q_\bot>0)$ as $|\mathbf{q}_\Vert|\to 0$, qualitatively altering the plasmon behavior and producing a small but finite gap \cite{greco2016,Hepting2022}. 
In this case, the long-wavelength limit of the density-density correlation function is approximately $\text{Re}\chi_{00}(\mathbf{q},\omega)\sim\left( \frac{|\mathbf{q}_\Vert|^2}{\omega^2}+\Delta \frac{q_\bot^2}{\omega^2}\right) $. Here, $\Delta=\left\langle v_\bot^2\right\rangle _F/\left\langle v_\Vert^2\right\rangle _F$ denotes the ratio of the Fermi surface averages of the squared  out-of-plane and in-plane velocities. The plasmon dispersion then reads 
$
	\omega_{pl}(|\mathbf{q}_\Vert|, q_\bot>0)\approx \hbar v_{pl}(q_\bot)\sqrt{|\mathbf{q}_\Vert|^2+\Delta q_\bot^2}
$.
Owing to the strong anisotropy of Sr$_2$RuO$_2$, the ratio $\Delta\sim 10^{-3}$ is small. We evaluated this ratio with the DFT-derived velocities since the local self-energy renormalizes the in-plane and out-of-plane components equally. Thus, over a wide region of energy-momentum space the plasmon dispersion may be accurately evaluated by neglecting the out-of-plane contribution (full colored lines in Fig.~4(a)). However, for momenta $|\mathbf{q}_\Vert| \lesssim  \sqrt{\Delta}\;q_\bot$, the dispersion starts to deviate from the linear behavior (dotted colored lines in Fig.~4(a)) and eventually becomes quadratic in $|\mathbf{q}_\Vert|$, 
$
\omega_{pl}(|\mathbf{q}_\Vert|\to0, q_\bot>0)\approx \tilde{\omega}_{pl}(q_\bot) \left[1+ \frac{|\mathbf{q}_\Vert|^2}{2\Delta q_\bot^2}\right]
$. The corresponding plasmon gap $\tilde{\omega}_{pl}(q_\bot)=\hbar v_{pl}(q_\bot)\sqrt{\Delta}\;q_\bot$ is of order $20$ meV.

\section{High-energy peak}

 \begin{figure}
	\includegraphics[width=0.65\textwidth]{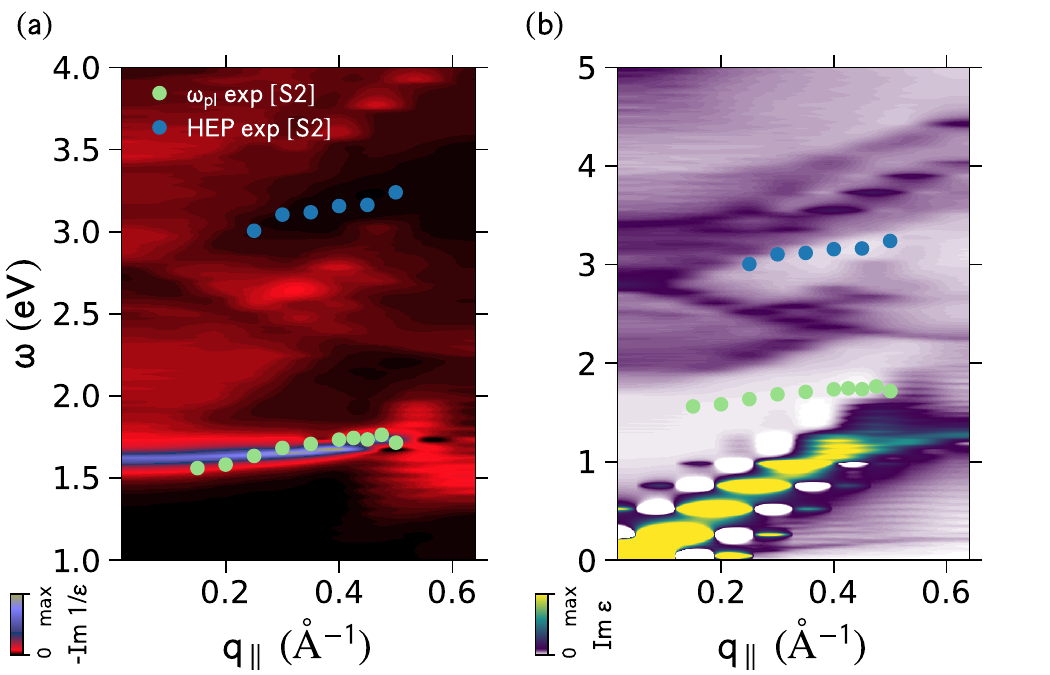}
	\renewcommand{\figurename}{Fig.}
	\caption{(a) EELS spectrum and (b) imaginary part of the dielectric function calculated within the full DFT. The full DFT can reasonably capture the measured optical plasmon dispersion $\omega_{pl}$ (green circles), but not the measured high-energy peak (HEP) (blue circles), which lies within the interband gap. The latter peak thus can be attributed to correlation effects. Experimental data is taken from Ref.~\cite{knupfer2022}.}
	\label{fig:eels_full_dft}
\end{figure}

In Fig.~\ref{fig:eels_full_dft}(a), we present the electron energy loss spectroscopy (EELS) spectrum calculated within the full DFT. As discussed in the previous section, it can describe well the optical plasmon energy. This is evident from Fig.~\ref{fig:eels_full_dft}(a), as the sharp maxima in the full DFT EELS spectrum agree with the plasmon peaks in the measured \cite{knupfer2022} EELS spectrum (green circles). However, the measured \cite{knupfer2022} EELS spectrum features another peak at higher energies (blue circles). Although full DFT EELS spectrum contains interband transitions beyond the three band model, it does not capture this high-energy peak (HEP); it is clearly evident that the HEP appears right in the energy-momentum space where the full DFT EELS spectrum has negligible spectral weight, that is, where the interband gap appears.  On other hand, the DFT+DMFT EELS spectrum in Fig.~2(a) of the main text nicely reproduces this HEP. This leaves us with the conclusion that the HEP is a direct signature of strong correlations. In particular, in Fig.~1(a) of the main text, we indicate possible transitions within the single-particle spectral function at the energy–momentum transfer at which the HEP is first observed in experiment; the associated features in the spectral function are driven by correlations.

\section{Low-energy peak}

 \begin{figure}
	\includegraphics[width=0.85\textwidth]{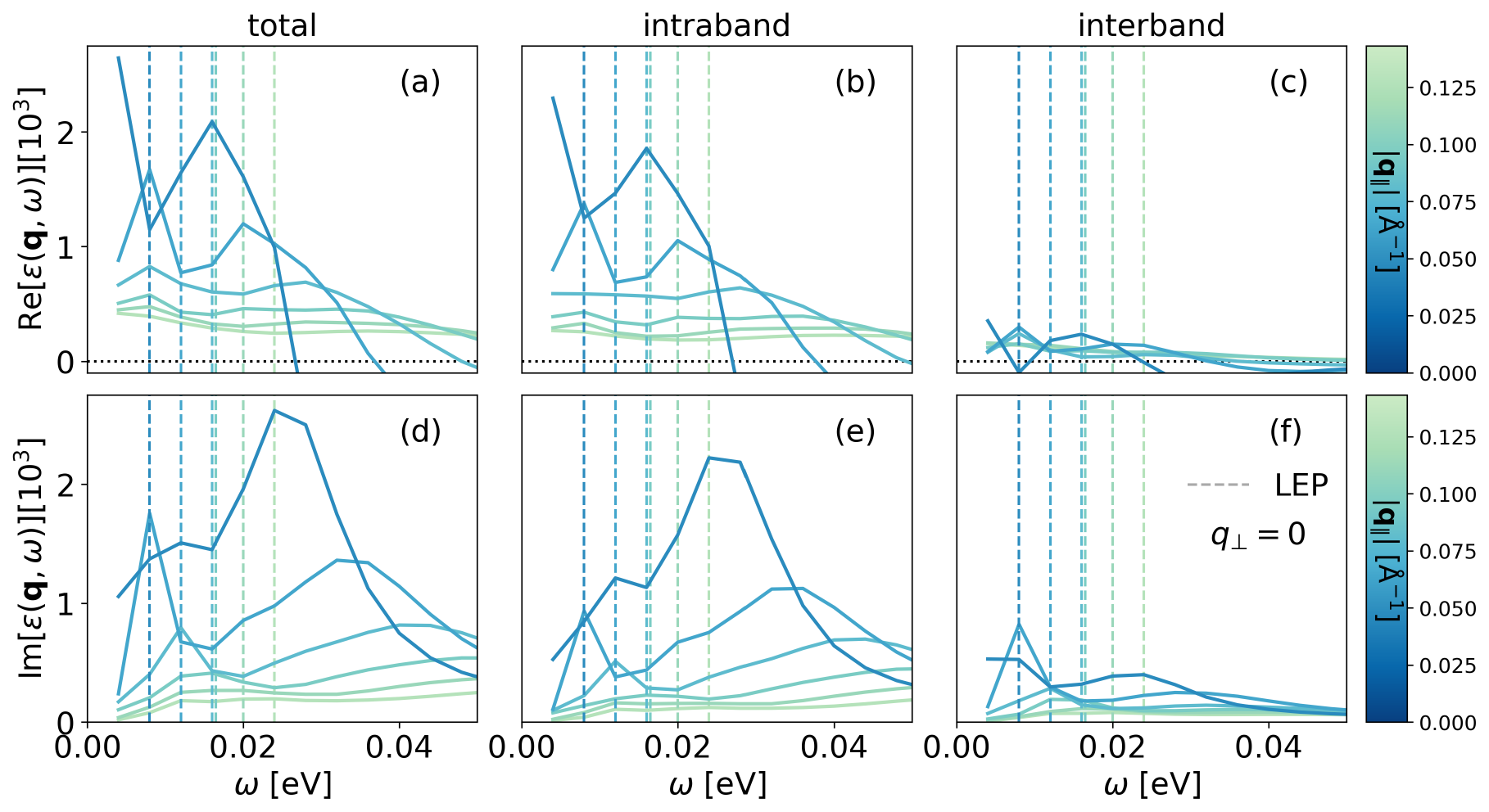}
	\renewcommand{\figurename}{Fig.}
	\caption{Real (a-c) and imaginary (d-f) parts of the DFT+DMFT dielectric function in the low-energy region for $q_\bot=0$ where the loss function exhibits a low-energy peak (LEP). The position of the LEP is indicated by dashed vertical lines.}
	\label{fig:intra_inter_low}
\end{figure}

In Fig.~4(c) of the main text, we show that the DFT+DMFT loss function exhibits a low-energy peak (LEP), whose positions, indicated by dashed vertical lines in Fig.~\ref{fig:intra_inter_low}, show reasonable agreement with the maxima observed in the measured EELS spectra that were attributed to the acoustic ``demon'' mode \cite{husain2023}. To gain deeper insight into this feature, Figs.~\ref{fig:intra_inter_low}(a-c) show the real part of the dielectric function, while Figs.~\ref{fig:intra_inter_low}(d-f) show its imaginary part in the relevant low-energy region for $q_\bot=0$. We further decompose the total dielectric function [Figs.~\ref{fig:intra_inter_low} (a,d)]—given by the sum over all $l,l',l'',l'''$ in Eq.~\eqref{eq:current-density-bubble}$-$into the intraband contribution [Figs.~\ref{fig:intra_inter_low} (b,e)], calculated by setting $l=l'=l''=l'''$, and the interband contribution [Figs.~\ref{fig:intra_inter_low} (c,f)], obtained as the remaining part. Note that the intraband contribution defined in this way may include both quasiparticle and incoherent features in the presence of strong correlations, with the latter more appropriately viewed as additional interband channels. While the real part$-$dominated by the intraband contribution$-$exhibits a shallow local minimum, it never crosses zero. Moreover, the damping is strong due to the large imaginary part. The LEP, though, occurs in the vicinity of the valley between two maxima of the imaginary part, emerging from the combined effect of intraband and interband contributions. Nevertheless, the LEP is strongly broadened and does not correspond to a genuine pole of the loss function.

 \begin{figure}
	\includegraphics[width=1\textwidth]{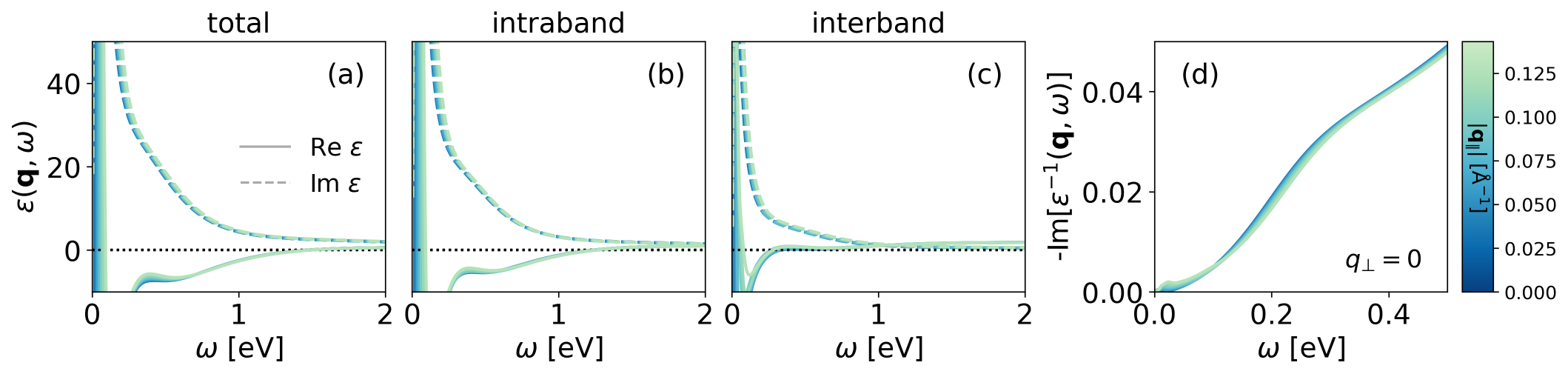}
	\renewcommand{\figurename}{Fig.}
	\caption{Real (solid lines) and imaginary (dashed lines) parts of the (a) total DFT+DMFT dielectric function for $q_\bot=0$ decomposed into (b) intraband and (c) interband contributions. The interband contribution exhibits a zero crossing around 0.3 eV, but it is outweighed by the intraband contribution, resulting in only a knee-like feature in the loss function (d).}
	\label{fig:intra_inter_high}
\end{figure}

 \begin{figure}
	\includegraphics[width=0.6\textwidth]{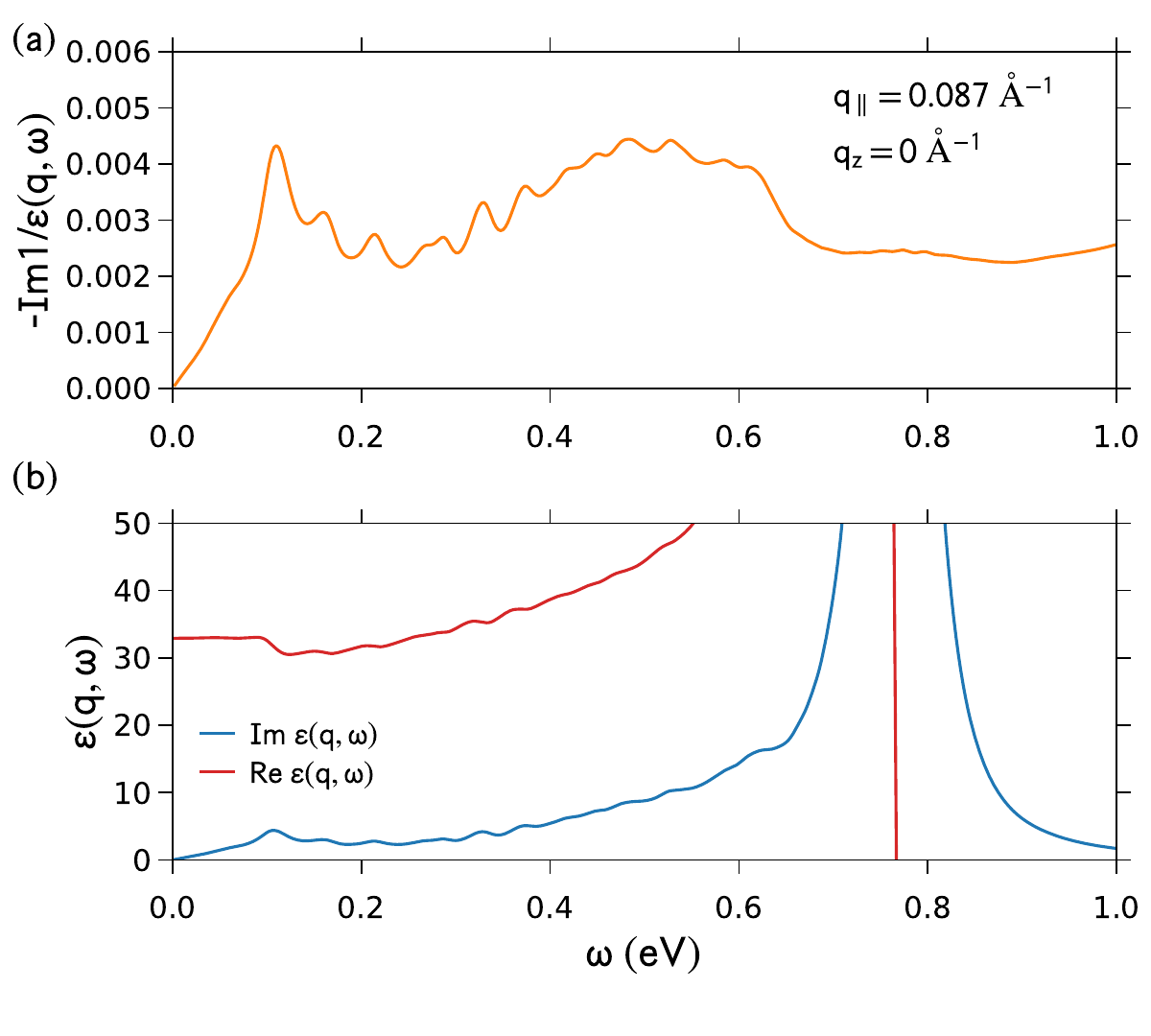}
	\renewcommand{\figurename}{Fig.}
	\caption{(a) Full DFT loss function and corresponding (b) real (red line) and imaginary (blue line) parts of the dielectric function. Similar to the DFT+DMFT case, the full DFT loss function exhibits a LEP and an additional, slightly higher-energy broad maximum reminiscent of the interband knee feature, though at shifted energies.}
	\label{fig:intra_inter_dft}
\end{figure}

We similarly decompose the total DFT+DMFT dielectric function for $q_\bot=0$ [Fig.~\ref{fig:intra_inter_high}(a)] into intraband [Fig.~\ref{fig:intra_inter_high}(b)] and interband [Fig.~\ref{fig:intra_inter_high}(c)] contributions in the larger energy range up to 2 eV. The interband part clearly crosses zero around 0.3 eV. If it were not for the intraband contribution, this zero crossing would correspond to a plasmon pole at that energy. However, the intraband contribution dominates, preventing a true pole; instead, a knee-like feature appears in the loss function, as shown in Fig.~\ref{fig:intra_inter_high}(d). Additionally, we observe that the intraband contribution alone would produce a plasmon at a lower energy than when the intraband and interband contributions are both included. This demonstrates that both intraband and interband contributions cooperate to generate the plasmons in Sr$_2$RuO$_4$, with the intraband part providing the dominant contribution.

Finally, we note from Fig.~\ref{fig:intra_inter_dft} that the LEP and a slightly higher-energy broad maximum reminiscent of the interband knee feature are also present in the full DFT calculation. Therefore, they are not uniquely correlation-driven effects, though renormalization shifts their positions. Note that 
the full DFT LEP likewise does not correspond to a well-defined pole in the loss function.

\bibliography{bibliography}